\documentclass[review,5p,fleqn]{elsarticle}

\usepackage{graphicx}
\usepackage{amsmath}
\usepackage{amsfonts}
\usepackage{amssymb}
\journal{Journal of Molecular Spectroscopy}

\begin{document}

\begin{frontmatter}



\title{A Global Algebraic Treatment for $XY_2$ Molecules : Application to $D_2S$.}

%%\author{}
%%\address{}
\author[Dijon,Tomsk]{O.V. Gromova}
\author[Dijon] {F. Michelot}
\author[Dijon] {C. Leroy \corref{cor}}
\cortext[cor]{Corresponding author.} \ead{claude.leroy@u-bourgogne.fr}
\author[Tomsk] {O.N. Ulenikov}
\author[Dijon] {Y. Pashayan-Leroy}
\author[Tomsk] {E.S. Bekhtereva}
\address[Dijon]{ Institut Carnot de Bourgogne, UMR 5209 CNRS-Universit\'{e} de Bourgogne, BP 47870, 21078 Dijon Cedex, France }
\address[Tomsk]{Laboratory of Molecular Spectroscopy, Physics Department,
Tomsk State University, 634050 Tomsk, Russia.}
\begin{abstract}
We suggest to use for $XY_2$ molecules some results  previously  established in a series of articles for vibrational modes and
electronic states with an $E$ symmetry type. We first summarize the formalism for the standard $u(2)\supset su(2)\supset so(2)$
chain which, for its most part, can be kept for the study of both stretching and bending modes of $XY_2$ molecules. Next the also
standard chain $u(3)\supset u(2) \supset su(2) \supset so(2)$ which is necessary, within the considered approach, is introduced
for the stretching modes. All operators acting within the irreducible representation (\textit{irrep}) $[N00]\equiv [N\dot{0}]$ of
$u(3)$ are built and their matrix elements computed within the standard basis. All stretch-bend interaction operators taking into
account the polyad structure associated with a resonance $\omega_1\approx \omega_3 \approx 2\, \omega_2$ are obtained. As an
illustration, an application to the $D_2S$ molecular system is considered, especially the symmetrization in $C_{2v}$. It is shown
that our unitary formalism allows to reproduce in an extremely satisfactory way all the experimental data up to the dissociation
limit.
\end{abstract}


\end{frontmatter}


\section{Introduction} \label{intro}
The hydrogen sulfide molecule and its isotopic species are of interest for terrestrial atmospheric pollutant measurements. As
this gas is more heavy than air, it remains concentrated to the floor level and can be lethal at high concentration. Global
warming process has increased the number of studies devoted to the chemical and physical properties of $H_2S$ and isotopic
species \cite{LAM2008,ZEL2008,FOR2007}. These molecules have also been observed in planet atmospheres like Jupiter \cite{FRI2005}
or Venus \cite{SMI2001} and appear
in the analysis of the interstellar medium \cite{MIN1989,HER1989,GRI1987}.\\
Many papers, devoted to the analysis of the rovibrational spectra of $D_2S$, have been published during the last thirty years :
anharmonicity corrections to observed fundamental frequencies of vibration in \cite{PLI1967}, various molecular structures have
been evaluated in \cite{COO1975}, \cite{GIL1985} for the $\nu_2$ band, \cite{CAM1988} for the three fundamental bands $\nu_2$,
$\nu_1$ and $\nu_3$ of $D_2{}^{32}S$ and the $\nu_2$ band of $D_2{}^{34}S$, application of the MORBID (Morse Oscillator Rigid
Render Internal Dynamics) computer program for the four isotopic molecules $H_2{}^{32}S$, $D_2{}^{32}S$ , $HD{}^{32}S$, and
$H_2{}^{34}S$ \cite{KOZ1994}.\\
Different theoretical models have been elaborated to improve the analysis of $D_2S$ \cite{LIU2006} or bent $XY_2$ molecules in
general \cite{HAL1987, IAC1990}. Initiated by the early works of Iachello and Oss \cite{IAC1990}, algebraic formalisms seem to be
good tools for the description of $XY_2$ molecules \cite{ZHE2000} particularly to take into account the local behavior. Also the
vibrational spectrum must be described using a Hamiltonian where the importance of Fermi-type interaction is taken into account
\cite{NAU2001}. Other studies using potential energy surface have been recently published \cite{TYU2001,TYU2004}. This last
method is particularly adapted when one has many data coming from different isotopic species. For further references about recent
analysis of $H_2S$ molecule and isotopes, the reader is invited to examine the references given
in \cite{LAM2008,TYU2001,TYU2004}.\\
In this paper, we use and adapt formalisms that we developed previously. Some of them dealt with unitary algebras applied in
molecular spectroscopy \cite{MIC1987,LER1992}, pure algebraic studies \cite{LER1994} or works concerning algebraic chains
which may be adapted to $XY_2$ systems \cite{MIC2004,MIC2005,MIC2007}.\\
In the theoretical part we use the properties of two algebraic chains, $u(2)\supset su(2)\supset so(2)$ and $u(3)\supset u(2)
\supset su(2) \supset so(2)$, to analyze the bending and stretching vibrational modes of bent $XY_2$ systems. Basis states and
operators oriented in these chains are built and the matrix elements of the oriented operators are given. Next symmetrization of
these tensors in $C_{2v}$ allows to build Hamiltonian and tensor operators adapted to the $C_{2v}$ molecular point group. A
method to select, in the $u(3) \otimes u(2)$ dynamical algebra, all relevant operators for a given polyad structure is
proposed and applied to the case of a 2:1 resonance.\\
As an illustration, in the last section our approach is tested upon the $D_2S$ molecule. The experimental data are reproduced
with a standard deviation close to $0.5 \mbox { cm}^{-1}$ and the calculated dissociation energy is found close to the
experimental one.
%********************************************************************
%*************Section 2************************************************
%********************************************************************
\section{Theoretical frame}\label{thf}
\subsection{The standard $u(2)\supset su(2)\supset so(2)$ chain} \label{sec21}
We consider first physical systems under the assumption that they can be described from two boson creation and annihilation
operators denoted $b_i^+$ and $b_i$ ($i=1,2$) which satisfy the usual Bose commutation relations. A basis for the space of states
may be obtained through repeated action of $b_1^+$ and $b_2^+$ onto the vacuum state $|0,0\rangle$:
\begin{equation}\label{eq201}
|n_1,n_2\rangle = (n_1!n_2!)^{-1/2} \ b_1^+{}^{n_1} \ b_2^+{}^{n_2} |0,0 \rangle .
\end{equation}
We note that these states may be taken as those of a two dimensional oscillator (within a usual approach) or as those of two one
dimensional oscillators. They will also be associated with the dynamical states for a one dimensional oscillator.\\
Tensor operators are built from the well-known Schwinger's realization of $su(2)$ in terms of two boson operators \cite{WYB1974}:
\begin{equation}\label{eq202}
\begin{array}{l}
J_+ =b_1^+b_2 ~~,~~J_- =b_2^+b_1 \\
J_z=  \frac{1}{2} (N_1-N_2) =\frac{1}{2} (b_1^+b_1-b_2^+b_2),
\end{array}
\end{equation}
with commutation relations
\[
\lbrack J_z,J_{\pm} \rbrack = \pm J_{\pm} \mbox{~~~,~~~} \lbrack J_+,J_- \rbrack = 2J_z ,
\]
and $u(2)$ is obtained with the addition of the linear invariant $N=N_1+N_2$, with $N_i \ |n_1,n_2\rangle = n_i \ |n_1,n_2\rangle
$ ($i=1,2$).
\subsubsection{General tensor operators and states within the standard chain} \label{sec211}
Keeping with previous conventions \cite{MIC2004} covariant $su(2)$ states $|j m\rangle\rangle$ and operators $T_{\,m}^{(j)}$ are
characterized by the relations ($m:-j,\cdots,j$):
\begin{equation}\label{eq2.1.1}
\begin{array}{l}
J_z \ |j m\rangle\rangle =  -m \ |j m\rangle\rangle\\
 J_{\pm} \ |j m\rangle\rangle =  -\lbrack (j\pm m)(j \mp m+1)\rbrack^{1/2}
\ |j m \mp 1\rangle\rangle ,
\end{array}
\end{equation}
and likewise for irreducible tensor operators (ITO)
\begin{equation}\label{eq2.1.2}
\begin{array}{l}
\lbrack J_z, \ T_{\,m}^{(j)} \rbrack  =  -m \ T_{\,m}^{(j)}\\
\lbrack J_{\pm},\; T_{\,m}^{(j)} \rbrack  =  -\lbrack (j\pm m)(j \mp m+1)\rbrack^{1/2} \ T_{m\mp 1}^{(j)}.
\end{array}
\end{equation}
In the whole $u(2)\supset su(2) \supset so(2)$ chain, symmetrized states and tensor operators are further characterized by an
additional $u(2)$ label $[m_{12}\ m_{22}]=[m_1\, -m_2]$ in Gel'fand notation \cite{GEL1950,LOU1970} with $j=(m_1+m_2)/2$ and
\begin{equation}\label{eq2.1.3}
\lbrack N_{i},{}^{[m_1\,-m_2]}T^{(j)}_{-j}\rbrack=m_{i2}{}^{[m_1\,-m_2]}T^{(j)}_{-j}~~i=1,2 .
\end{equation}
With equations (\ref{eq202}, \ref{eq2.1.2}) it can be checked that a realization for the extremal components appearing in
(\ref{eq2.1.3}) is given by:
\begin{equation}\label{eq2.1.4}
\mbox{{\Large T}} \left\lbrack \!
\begin{array}{c}
m_1 \  -m_2 \\
m_1
\end{array} \!
\right\rbrack = (-1)^{m_2} \, i^{m_1} \lbrack m_1!\,m_2! \rbrack^{-\frac{1}{2}} \, b_1^+{}^{m_1}b_2{}^{m_2}
\end{equation}
This allows to generate two sets of fundamental operators
\begin{equation}\label{eq2.1.5}
\begin{split}
^{[m_1\,0]}T^{(j)}_{~m} &= (-1)^{m_1+m} \,[(\frac{m_1}{2}-m)!(\frac{m_1}{2}+m)!]^{-1/2}   \\ &\times \,
b_1^+{}^{(\frac{m_1}{2}-m)} \, b_2^+{}^{(\frac{m_1}{2}+m)},
\end{split}
\end{equation}
\begin{equation}\label{eq2.1.6}
\begin{split}
^{[0\,-m_2]}T^{(j)}_{~m} &= (-1)^{m_2}\,[(\frac{m_2}{2}-m)!(\frac{m_2}{2}+m)!]^{-1/2} \\
&\times \, b_1{}^{(\frac{m_2}{2}+m)} \,  b_2{}^{(\frac{m_2}{2}-m)},
\end{split}
\end{equation}
with $j=m_2/2$. In particular the standard covariant basis is obtained with (\ref{eq2.1.5}) acting upon the vacuum state:
\begin{equation}\label{eq2.1.7}
|[n\,0]jm\rangle\rangle \equiv {}^{[n\,0]}T^{(j)}_{~m} \ |0,0> = (-1)^{2j+m} |j-m,j+m \rangle \rangle.
\end{equation}
>From the previous set (\ref{eq2.1.5}, \ref{eq2.1.6}) one may build all functionally independent operators which may act within
the \textit{irrep} $[n\,0]$ of $u(2)$ through
\begin{equation}\label{eq2.1.8}
^{[m_1\,-m_2]}\mathcal{T}^{(j)}_{~m} = i^{j-j_{max}}\! \left\lbrack ^{[m_1\,0]}T^{(\frac{m_1}{2})}\!\times
^{[0\,-m_2]}T^{(\frac{m_2}{2})} \!\right\rbrack^{(j)}_{m}
\end{equation}
with
\[
j_{min}=\frac{|m_1-m_2|}{2} \leq j \leq \frac{m_1+m_2}{2}=j_{max} .
\]
Their expansion in normal ordered form can easily be obtained; alternatively they may be written
\begin{equation}\label{eq2.1.9}
^{[m_1\,-m_2]}\mathcal{T}^{(j)}_{~m} = {}^{\lbrace m_1\,m_2\rbrace} \mbox{{\Large g}}_j(N_1\!+\!N_2) \ \
^{[m_1'\,-m_2']}T^{(j)}_{~m},
\end{equation}
where
\begin{equation}\label{eq2.1.10}
\begin{array}{lcr}
m_1'  =  \frac{m_1-m_2}{2}+j  & ,& m_2'  =  \frac{m_2-m_1}{2}+j,
\end{array}
\end{equation}
hence $j=(m_1'+m_2')/2$. ${}^{\lbrace m_1\,m_2\rbrace}\mbox{{\Large g}}_j$ is a polynomial function of the $u(2)$ linear
invariant $N_1+N_2$ given by:
\begin{eqnarray}\label{eq2.1.11}
\lefteqn{{}^{\lbrace m_1\,m_2\rbrace}\mbox{{\Large g}}_j(N_1\!+\!N_2) = \left\lbrack \frac{(2j+1)!}{(j_{max}+j+1)!\,(j_{max}-j)!}
\right\rbrack^{\frac{1}{2}} }\nonumber \\
&&\times\,(N_1+N_2+\frac{m_2-m_1}{2}-j)^{\lbrack j_{max}-j\rbrack},
\end{eqnarray}
where $X^{\lbrack k\rbrack}=X\times (X-1)\times...\times(X-k+1)$.\\
We note that when $j=j_{max}$, the operator (\ref{eq2.1.11}) reduces to the identity and
\begin{equation}\label{eq2.1.11a}
^{[m_1\,-m_2]}\mathcal{T}^{(j_{max})}\equiv {}^{[m_1\,-m_2]}T^{(j_{max})},
\end{equation}
the minimal covariant component of which is given by (\ref{eq2.1.4}). All phase conventions have been settled so that under
hermitian conjugation ($\dag$) and time reversal ($\mathcal{K}_t$) we have
\begin{equation}\label{eq2.1.12}
\begin{split}
^{[m_1\,-m_2]}\mathcal{T}^{\dagger(j)}_{~~m} &= (-1)^{j-m} \ \left( ^{[m_1\,-m_2]}\mathcal{T}^{(j)}_{-m} \right)^{\dagger},
\\&= i^{m_1-m_2} \ ^{[m_2\,-m_1]}\mathcal{T}^{(j)}_{~m},
\end{split}
\end{equation}
\begin{equation}\label{eq2.1.13}
\mathcal K_t \ ^{[m_1\,-m_2]}\mathcal{T}^{(j)}_{~m} \ \left. \mathcal K_t\right.^{-1}
=(-1)^{m'_1}{}^{[m_1\,-m_2]}\mathcal{T}^{(j)}_{m}.
\end{equation}
\subsubsection{Matrix elements within the standard basis} \label{sec212}
With the Wigner-Eckart's theorem we have in terms of $su(2)$ Clebsch-Gordan coefficients (CG) \cite{EDM1974}:
\begin{eqnarray}\label{eq2.1.14}
\lefteqn{ \langle\langle[n''\,0]j''m'' |{} ^{[m_1\,-m_2]}\mathcal{T}^{(j)}_{~m}{}|[n'\,0]j'm' \rangle\rangle
=(2j''+1)^{-1/2} } \nonumber \\
 & &\times C\begin{array}{ccc} m & m' &(j'')\ast \\
                   (j & j')& m''\end{array}
\left([n''\,0]j''||{}^{[m_1\,-m_2]}\mathcal{T}^{(j)}{}||[n'\,0]j'\right)\nonumber\\
& &=F\begin{array}{ccc} m & m' & ([n''\,0]j'')\ast \\
([m'_{1} -m'_{2}]j & [n'\,0]j' ) & m'' \end{array} \nonumber \\
&& \times (2j''+1)^{-1/2}\left([n''\,0]j''||{}^{[m_1\,-m_2]}\mathcal{T}^{(j)}{}||[n'\,0]j'\right),
\end{eqnarray}
where $\ast$ denotes complex conjugation. The notation for the $F$ symbols, which retains the full $u(2) \supset su(2)$ labels,
is useful when symmetry adaptation in a point group is performed. Reduced matrix elements (\textit{rme}) for all operators are
obtained with
\begin{flalign}\label{eq2.1.15}
\left([n''\,0]j''||{}^{[m_1\,-m_2]}\mathcal{T}^{(j)}{}||[n'\,0]j'\right)
= \delta_{n'',n'+m_1-m_2}  \nonumber \\
 \times  \frac{i^{-m_2'}}{(n'-m_2)!}\left\lbrack \!\frac{(2j+1)(n'+m_1'+1)!\,(n'-m_2')!}{(j_{max}+j+1)!\,(j_{max}-j)!}
\right\rbrack^{\frac{1}{2}}.
\end{flalign}
\subsection{ Tensors adapted to an $u(3)$ dynamical algebra} \label{sec22}
For the applications we have in mind, the initial assumptions are the following:\\
$\bullet$ A molecule, with point group symmetry $G$ ($G=C_{2v}$), admits in its full vibrational representation two non
degenerate modes with close enough frequencies.\\
$\bullet$ The interaction of these modes with other vibrational modes is sufficiently low so that in first approximation it can
be neglected. As a consequence a separate study taking into account the degrees of freedom associated with these modes only is
possible.\\
$\bullet$ This study, made within the frame of an $u(p+1)$ dynamical approach, requires the introduction of a dynamical or
non-invariance $u(3)$ algebra to which we associate the elementary boson operators $\lbrace b_i^+,b_i\rbrace_{i=1,2,3}$.
\subsubsection{The algebraic chain $u(3)\supset u(2)\supset su(2)\supset so(2)$} \label{sec221}
>From the preceding assumptions the space of states is a carrier space for the so-called totally symmetric (or most degenerate)
\textit{irrep} $[N00]= [N\,\dot{0}]$ of $u(3)$ which subduces to $[n\,0]$ ($n=0, 1, \cdots N$) in $u(2)$ \cite{LOU1970,HOL1971}.
Then all operators which may act within this \textit{irrep} are of symmetry $\lbrack
z,0,-z \rbrack$ with $z=0,1,\ldots,N$ (see \cite{MIC2004,LER1991,BOU1996} for more details).\\
Those which are maximal in  $u(3)$ have the form:
\begin{equation}\label{eq2.2.1}
\mbox{{\Large T}} \left\lbrack
\begin{array}{c}
\lbrack z~ 0 -z \rbrack \\
(max_c)
\end{array}
\right\rbrack  = \left\lbrack \frac{(N-z)!}{N!\; z!}\right\rbrack^{1/2} \ b_1^+{}^z b_3{}^z,
\end{equation}
where the $\alpha_{z,N}$ coefficient is determined through the following normalization condition:
\begin{equation}\label{eq2.2.2}
<n,0,N-n| \ \mbox{{\Large T}} \left\lbrack
\begin{array}{c}
\lbrack z~ 0 -z \rbrack \\
(max_c)
\end{array}
\right\rbrack \ |0,0,N> = \delta_{z,n}.
\end{equation}
The notation $|n_1, n_2, n_3 \rangle$ for the states is that of the $u(3)$ canonical chain. $|0, 0, N \rangle$ and $|n, 0, N-n
\rangle$ represent respectively the state with zero excitation quantum and the state with $n$ excitation quanta maximal in
$u(2)$. Then the semi-maximal operators of the $u(3)$ dynamical algebra write as:
\begin{equation}\label{eq2.2.3}
\mbox{{\Large T}} {\renewcommand{\arraystretch}{1}
\renewcommand{\arraycolsep}{-0.05cm}
\left\lbrack
\begin{array}{ccccc}
\, z \,\, & & \, 0 \,\, & & \! -z \, \ \\
& m_1 & & \!-m_2 & \\
& & m_1 & &
\end{array}
\right\rbrack} = \mathcal G(z,m_1,m_2) \ b_1^+{}^{m_1} \ b_2{}^{m_2} ,
\end{equation}
where $\mathcal{ G}(z,m_1,m_2)$  is an operator valued function invariant in $u(2)$ and defined by:
\begin{eqnarray}\label{eq2.2.4}
\lefteqn{\mathcal G(z,m_1,m_2)=  \left\lbrack \binom{z}{m_1} \binom{z}{m_2} \right. } \nonumber \\
&\times \displaystyle{\left. \frac{(z+m_2+1)!\, (z+m_1+1)!\,(N-z)!}{(m_1+m_2+1)!\,(2z+1)!\,N!\,z!} \right\rbrack^{\frac{1}{2}}}
\nonumber \\
&\times\left\lbrace \sum\limits_{t=0}^{u} \displaystyle{(-1)^{t+m_2} \binom{z-m_1}{t} \frac{(m_1+m_2+1)!}{ (m_1+m_2+1+t)!}}
\right. \nonumber \\ &\!\times \!\displaystyle{\left.\frac{(z-m_2)!}{(z-m_2-t)!} (N_1+N_2-m_1)^{\lbrack t \rbrack}
\right.}\nonumber\\
&\times \displaystyle{\left. (N_3-z+m_1+u)^{\lbrack u-t \rbrack} \right\rbrace   b_3^+{}^{z-m_1-u} \ b_3{}^{z-m_2-u} }.
\end{eqnarray}
with $u=inf(z-m_1,z-m_2)$. From equations (\ref{eq2.1.4}, \ref{eq2.1.11a}) it appears the left member of equation (\ref{eq2.2.3})
is, within a phase factor, the minimal covariant component of an ITO within the $su(2) \supset so(2)$ chain. So from the results
in section \ref{sec211}, and taking into account that $\mathcal {G}_(z,m_1,m_2)$ commutes with the $su(2)$ ladder operators $J_+$
and $J_-$ , an arbitrary covariant operator is obtained through:
\begin{eqnarray}\label{eq2.2.5}
\mbox{{\Large T}} {\renewcommand{\arraystretch}{1} {\renewcommand{\arraycolsep}{-0.05cm} \left\lbrack
\begin{array}{c}
z\ \ \ 0 \ \ \ -z \\ \lbrack m_1\,  -m_2 \rbrack \ (j) \\
m
\end{array}
\right\rbrack}} =(-1)^{m_2}[m_1!m_2!]^{1/2}  \nonumber \\
~~ \times \, \mathcal G(z,m_1,m_2) \ ^{[m_1\,-m_2]}T^{(j)}_{~m},
\end{eqnarray}
where the various labels may take the values $0 \leq z \leq N$,  $ 0 \leq m_1 \leq z$,  $ 0 \leq m_2 \leq z$, $j=(m_1+m_2)/2$, $
-j\leq m \leq j$. The phase of the preceding operators have been chosen so that
\begin{eqnarray}\label{eq2.2.6}
 \mbox{{\Large T}} {\renewcommand{\arraystretch}{1}
{\renewcommand{\arraycolsep}{-0.05cm} \left\lbrack
\begin{array}{c}
z\ \ \ 0 \ \ \ -z \\ \lbrack m_1 \, -m_2 \rbrack \ (j) \\
m
\end{array}
\right\rbrack}}^{\dagger} = (-1)^{j+m} \ i^{m_1-m_2} \nonumber\\
~~\times \mbox{{\Large T}} {\renewcommand{\arraystretch}{1} {\renewcommand{\arraycolsep}{-0.05cm} \left\lbrack
\begin{array}{c}
z\ \ \ 0 \ \ \ -z \\ \lbrack m_2 \, -m_1 \rbrack \ (j) \\
-m
\end{array}
\right\rbrack}},
\end{eqnarray}
\begin{eqnarray}\label{eq2.2.7}
\mathcal K_t \, \mbox{{\Large T}} {\renewcommand{\arraystretch}{1} {\renewcommand{\arraycolsep}{-0.05cm} \left\lbrack
\begin{array}{c}
z\ \ \ 0 \ \ \ -z \\ \lbrack m_1 \, -m_2 \rbrack \ (j) \\
m
\end{array}
\right\rbrack}} \mathcal K^{-1}_t = (-1)^{m_1} \nonumber\\ ~~~~ \times \mbox{{\Large T}} {\renewcommand{\arraystretch}{1}
{\renewcommand{\arraycolsep}{-0.05cm} \left\lbrack
\begin{array}{c}
z\ \ \ 0 \ \ \ -z \\ \lbrack m_1 \, -m_2 \rbrack \ (j) \\
m
\end{array}
\right\rbrack}}.
\end{eqnarray}
\subsubsection{Matrix elements}\label{sec222}
The covariant states, associated with the representation $\lbrack N\,\dot{0} \rbrack$ of $u(3)$ and adapted to the subduction
$u(3)\supset u(2)\supset su(2)\supset so(2)$, are denoted $| [N\, \dot{0}]\lbrack n\,0\rbrack j m \rangle\rangle$ in the
following. They may be generated from the zero quantum excitation state $| [N\, \dot{0}]\lbrack 0\,0\rbrack 0 0 \rangle\rangle
\equiv |0,0,N>$, through the relation:
\begin{equation}\label{eq2.2.8}
| [N\, \dot{0}]\lbrack n\,0\rbrack j m \rangle\rangle= \mbox{{\Large T}} {\renewcommand{\arraystretch}{1}
{\renewcommand{\arraycolsep}{-0.05cm} \left\lbrack
\begin{array}{c}
n \ 0 \ -n \\ \lbrack n \ 0 \rbrack \ (j) \\
m
\end{array}
\right\rbrack}}\ | [N\,\dot{0}] \lbrack 0\,0\rbrack 0 0 \rangle\rangle.
\end{equation}
The matrix elements for operators (\ref{eq2.2.5}), computed in basis (\ref{eq2.2.8}), are then given by:
\begin{eqnarray}\label{eq2.2.9}
\lefteqn{ \langle\langle[N\,\dot{0}]\,[n''\,0]j''m'' |{} \mbox{{\Large T}} {\renewcommand{\arraystretch}{1}
{\renewcommand{\arraycolsep}{-0.05cm} \left\lbrack
\begin{array}{c}
z\ \ \ 0 \ \ \ -z  \\ \lbrack m_1 \, -m_2 \rbrack \ (j) \\
m
\end{array}
\right\rbrack}}{}|[N\,\dot{0}]\,[n'\,0]j'm' \rangle\rangle } \nonumber
\\ & &= (2j''+1)^{-1/2}
C\begin{array}{ccc} m & m' &(j'')\ast \\
                   (j & j')& m''\end{array} \\
& & \times \left([N\dot{0}]\,[n''\,0]j''||{}\mbox{{\Large T}} {\renewcommand{\arraystretch}{1}
{\renewcommand{\arraycolsep}{-0.05cm} \left\lbrack
\begin{array}{c}
z \ 0 \ -z \\ \lbrack m_1 \, -m_2 \rbrack \ (j)
\end{array}
\right\rbrack}}{}||[N\,\dot{0}]\,[n'\,0]j'\right). \nonumber
\end{eqnarray}
Using the expanded form (\ref{eq2.2.4}) of $\mathcal{ G}(z,m_1,m_2)$ we obtain the \textit{rme} in the form
\begin{eqnarray}\label{eq2.2.10}
\lefteqn{ ( [N\,\dot{0}]\,[n''\,0]j''  || \mbox{{\Large T}}{\renewcommand{\arraystretch}{1} {\renewcommand{\arraycolsep}{-0.05cm}
\left\lbrack
\begin{array}{c}
z\ \ \ 0 \ \ \ -z  \\ \lbrack m_1  -m_2 \rbrack \ (j)
\end{array}
\right\rbrack}}{}  || [N\,\dot{0}]\,[n'\,0]j') =} \nonumber
\\  & \nonumber \\ &
\displaystyle{ \left\lbrack \binom{z}{m_1} \binom{z}{m_2} \
\frac{(z+m_2+1)!\,(z+m_1+1)!\,(N-z)!}{(m_1+m_2+1)!\,(2z+1)!\,N!\,z!}\right.}
\nonumber\\
 & \displaystyle{\left. \times \frac{m_1!\,m_2!\,(N-n')!\,(N-n'+m_2-m_1)!}{ (N-n'-z+m_2+u)!\,(N-n'-z+m_2+u)!} \right\rbrack^{1/2}} \nonumber \\  &
\nonumber \\  & \times \displaystyle{\left\lbrace \sum\limits_{t}^{u} (-1)^{t} \binom{z-m_1}{t}
\frac{(m_1+m_2+1)!(z-m_2)!}{(m_1+m_2+1+t)!}\right. }\nonumber\\ &\! \! \displaystyle{\left. \times
\frac{(n'-m_2)!(N-n'+m_2-z+u)!}{(z-m_2-t)!(n'-m_2-t)! (N-n'+m_2-z+t)!} \right\rbrace }\nonumber  \\  & \times
([N\,\dot{0}]\,[n''\,0]j'' \ ||{}^{[m_1\,-m_2]}T^{(j)}|| \,[n'\,0]j'),
\end{eqnarray}
where the remaining \textit{rme} is given by (\ref{eq2.1.15}) with $m'_1=m_1$, $m'_2=m_2$.
%********************************************************************
%*************Section 3************************************************
%********************************************************************
\section{Application to $XY_2$ molecules}\label{appxy}
We make the assumption that an appropriate dynamical algebra for the description of the vibrational spectrum of these molecules
is $u(3)_S \times u(2)_B$, where the indices $S$ and $B$ respectively refer to stretching and bending modes. Below, we first
specify the notations which are used in for each type of mode and next perform the symmetry adaptation in $C_{2v}$ for states and
operators.
\subsection{Bending mode $\nu_2$}\label{bend}
\subsubsection{Standard states and operators}\label{bend1}
For the Schwinger's realization of $su(2)_B$ we take (\ref{eq202}):
\[
\begin{array}{l}
{}^{(b)}J_+ = b_4^+b_5 ~~,~~{}^{(b)}J_- = b_5^+b_4 \\
{}^{(b)}J_z=  \frac{1}{2} (N_4-N_5) =\frac{1}{2} (b_4^+b_4-b_5^+b_5),
\end{array}
\]
and  $^{(b)}I_1^{(2)}=N_4+N_5=\widehat{N}_b$ for the $u(2)_B$ linear invariant. The standard covariant basis (\ref{eq2.1.7}) is
written ($N_b=2J_b$):
\begin{flalign}\label{eqb.2}
|[N_b\,0]J_b\,m \rangle\rangle  =   (-1)^{2J_b+m}\ |J_b-m,J_b+m> \nonumber \\
 =  \frac{(-1)^{2J_b+m} }{\lbrack (J_b+m)!\,(J_b-m)! \rbrack^{1/2}}\,  b_4^+{}^{J_b-m} b_5^+{}^{J_b+m} \ |0,0>.
\end{flalign}
We may also set
\begin{equation*}\label{eqb.3}
\begin{array}{lcl}
J_b-m=n_4=v_2& & J_b=\displaystyle{\frac{n_4+n_5}{2}=\frac{N_b}{2}},\\
&\Leftrightarrow&\\ J_b+m=n_5=N_b-v_2&&m=\displaystyle{\frac{n_5-n_4}{2}=\frac{N_b}{2}-v_2}.
\end{array}
\end{equation*}
Thus all results of section \ref{sec21} can be used with the appropriate change of indices. However as $N_b$ is associated with
the maximal number of bending states and that only operators acting within the $[N_b\,0]$ \textit{irrep} of $u(2)$ are allowed we
necessarily have $m_4=m_5$ for the general operators built before. The vibrational operators are thus ($0 \leq j_b \leq m_4$)
\begin{flalign}\label{eqb.7b}
^{[m_4\,-m_4]}\mathcal{B}^{(j_b)}_{~m} &= i^{j_b-m_4} \left\lbrack {}^{[m_4\,0]}T^{(\frac{m_4}{2})}\times
{}^{[0\,-m_4]}T^{(\frac{m_4}{2})} \right\rbrack^{(j_b)}_{m} \nonumber \\
&={}^{\lbrace m_4,m_4\rbrace} \mbox{{\Large g}}_{j_b}(N_4\!+\!N_5) \ \ ^{[m'_4\,-m'_4]}B^{(j_b)}_{~m},
\end{flalign}
with $m'_4=m'_5=j_b$ and
\begin{flalign}\label{eqb.11}
^{\lbrace m_4,m_4\rbrace}\mbox{{\Large g}}_{j_b}(N_4\!+\!N_5) &= \left\lbrack \frac{(2j_b+1)!}{(m_4+j_b+1)!\,(m_4-j_b)!}
\right\rbrack^{1/2} \nonumber \\ & \times (N_4+N_5-j_b)^{\lbrack m_4-j_b\rbrack}.
\end{flalign}
Their expanded expression in normal ordered form is:
\begin{eqnarray}\label{eqb.8}
\lefteqn{^{[m_4\,-m_4]}\mathcal{B}^{(j_b)}_{~m} = i^{j_b-m_4}\!\sum\limits_{q_1,q_2} i^{2 q_1}\! \left\lbrack \left(
\frac{m_4}{2}-q_1
\right)! \left( \frac{m_4}{2}+q_1 \right)!\right. }\nonumber \\
 & \times \displaystyle{\left.   \left( \frac{m_4}{2}-q_2 \right)!
\left( \frac{m_4}{2}+q_2 \right)! \right\rbrack^{-1/2} {}C\begin{array}{ccc}q_1&q_2&(j_b)\\(\frac{m_4}{2}&\frac{m_4}{2})&m
\end{array}}\nonumber \\
& \times \,\displaystyle{ b_4^+{}^{\frac{m_4}{2}-q_1} \, b_5^+{}^{\frac{m_4}{2}+q_1} \, b_4{}^{\frac{m_4}{2}+q_2} \,
b_5{}^{\frac{m_4}{2}-q_2}}.
\end{eqnarray}
With equations (\ref{eq2.1.14}, \ref{eq2.1.15}) their matrix elements are
\begin{eqnarray}\label{eqb.13}
\lefteqn{ \langle\langle[N_b\,0]J_b\, m'' |{} ^{[m_4\,-m_4]}\mathcal{B}^{(j_b)}_{~m}{}|[N_b\,0]J_b\, m' \rangle\rangle
=(2J_b+1)^{-\frac{1}{2}} }\nonumber \\
&& \times F\begin{array}{ccc} m & m' & ([N_b\,0]J_b)\ast \\
([m'_{4} -m'_{4}]j_b & [N_b\,0]J_b ) & m'' \end{array} \nonumber \\
&&\times \left([N_b\,0]J_b||{}^{[m_4\,-m_4]}\mathcal{B}^{(j_b)}{}||[N_b\,0]J_b\right),
\end{eqnarray}
with for the \textit{rme}
\begin{flalign}\label{eqb.14b}
\left([N_b\,0]J_b||{}^{[m_4\,-m_4]}\mathcal{B}^{(j_b)}{}||[N_b\,0]J_b\right) =\frac{i^{-j_b}}{(N_b-m_4)!} \nonumber \\ \times
\left\lbrack \frac{(2j_b+1)(N_b+j_b+1)!\,(N_b-j_b)!}{(m_4+j_b+1)!\,(m_4-j_b)!} \right\rbrack^{\frac{1}{2}}.
\end{flalign}
\subsubsection{Symmetrization in $C_{2v}$}\label{bend2}
For the $\nu_2$ mode with $A_1$ symmetry there is \textit{a priori} no difficulty. All operators
$^{[m_4\,-m_4]}\mathcal{B}^{(j_b)}_{~m}$ are of $A_1$ species as well as any linear combination
\begin{equation}\label{eq3.2.6}
 ^{[m_4\,-m_4]}\mathcal{B}^{(j_b)}_{rA_1} = \sum\limits_m {~}^{[m_{4}\,-m_{4}]} G_{rA_1}^{m} \
 ^{[m_4\,-m_4]}\mathcal{B}^{(j_b)}_{~m},
\end{equation}
where ${~}^{[m_{4}\,-m_{4}]} G$ is a unitary matrix and $r=1, \cdots ,2j_b+1$ a multiplicity index. However a consistent tensor
formalism imposes some restrictions upon the $G$ matrix.\\
The matrices for the \textit{irrep} of $C_{2v}$, which reduce to characters, being real, the metric tensor may be chosen (and is
usually chosen) identical to the identity. This traduces by:
\[
T^{\dagger(\Gamma)}_{\ \sigma}= T^{(\Gamma)\dagger}_{~\sigma} ,
\]
for any symmetrized operator, and in particular for the elementary ones
\[
b^{(A_1)\dagger}=b^{\dagger(A_1)}=b^{+(A_1)} .
\]
For our problem we will not undertake a general discussion as was made for $E$ modes \cite{MIC2004}. For $\nu_2$ we only need the
similarity transformations ${~}^{[N_b\,0]} G$ for the states and ${~}^{[m_{4}\,-m_{4}]} G$ for the operators. Also from the
expression (\ref{eqb.11}) for ${}^{\lbrace m_4,m_4\rbrace}\mbox{{\Large g}}_{j_b}(N_4\!+\!N_5)$, with $A_1$ symmetry and
invariant upon time reversal, we see that we only need to determine ${~}^{[m_{4}\,-m_{4}]} G$ for the operators
$^{[m_4\,-m_4]}B^{(j_b)}$ ($j_b=m_4$). That is (equation (\ref{eqb.7b})):
\begin{equation}\label{eqb.19}
\begin{split}
 ^{[m_4\,-m_4]}\mathcal{B}^{(j_b)}_{rA_1} &= {}^{\lbrace m_4,m_4\rbrace}\mbox{{\Large g}}_{j_b}(N_4\!+\!N_5)  \\
&\times \sum\limits_m {~}^{[m'_{4}\,-m'_{4}]} G_{rA_1}^{m} \ ^{[m'_4\,-m'_4]}B^{(j_b)}_{~m}.
\end{split}
\end{equation}
$\bullet$ \textit{Symmetrized states}. From equations (\ref{eq2.1.7},\ref{eq2.1.13}) we have
\begin{equation}\label{eqb.20}
\begin{split}
\mathcal K_t|[N_b\,0]J_b\,m\rangle\rangle &\equiv \mathcal K_t{}^{[N_b\,0]}B^{(J_b)}_{~m}\mathcal K_t^{-1} \mathcal K_t\ |0,0> \\
& = (-1)^{N_b}|[N_b\,0]J_b\,m\rangle\rangle .
\end{split}
\end{equation}
If we set
\[
|[N_b\,0]J_b\,rA_1\rangle\rangle =e^{i\theta}|[N_b\,0]J_b\,m\rangle\rangle ,
\]
we get symmetrized states invariant upon time reversal if $e^{2i\theta}=(-1)^{N_b}= i^{2N_b}$, thus a possible choice is:
\begin{equation}\label{eqb.21}
\begin{split}
|[N_b\,0]J_b\,rA_1\rangle\rangle \equiv |[N_b\,0]J_b\,m A_1\rangle\rangle =i^{N_b}|[N_b\,0]J_b\,m\rangle\rangle  \\
\equiv \, (-1)^{J_b-m}\ |n_4=J_b-m,n_5=J_b+m>,
\end{split}
\end{equation}
which amounts to choose
\begin{equation}\label{eqb.22}
 {~}^{[m_{4}\,0]} G_{rA_1}^{m}=i^{m_4} \delta_{r,m}.
\end{equation}
$\bullet$ \textit{Symmetrized operators}. Likewise from the properties (\ref{eq2.1.12}, \ref{eq2.1.13}) under hermitian
conjugation and time reversal of the standard tensors, it may be shown that we may built the hermitian symmetrized operators
\begin{equation}\label{eqb.30}
\begin{split}
^{[m_4\,-m_4]}\mathcal{B}^{(j_b)}_{|m|{\varepsilon}A_1}&=\frac{i^{\varepsilon}}{\sqrt{2}}
\left(^{[m_4\,-m_4]}\mathcal{B}^{(j_b)}_{~m}\right. \\
&\left.+(-1)^{\varepsilon}(-1)^{j_b+m}\,^{[m_4\,-m_4]}\mathcal{B}^{(j_b)}_{-m}\right),  \\
^{[m_4\,-m_4]}\mathcal{B}^{(j_b)}_{0A_1}&=i^{m_4}\,^{[m_4\,-m_4]}\mathcal{B}^{(j_b)}_{~0},
\end{split}
\end{equation}
with $\varepsilon=0,1$. Operators characterized by $\varepsilon=0$ are invariant upon time reversal (resp. non invariant upon
time reversal) for $j_b$ even (resp. $j_b$ odd); it is the reverse for those characterized by $\varepsilon=1$. Operators
``diagonal in $v_2$'' $^{[m_4\,-m_4]}\mathcal{B}^{(j_b)}_{0A_1}$ are all invariant upon time reversal. We may thus write an
effective bending Hamiltonian:
\begin{equation} \label{eqb.39}
\mathcal{H}_B=\displaystyle{\sum_{m_4=0}^{N_b}}\displaystyle{\sum_{j_b=0}^{m_4}}\,^{\{m_4\}}t_{b}^{(j_b)}\,{}^{[m_4\,-m_4]}\mathcal{B}^{(j_b)}_{0
A_1}.
\end{equation}
$\bullet$ \textit{Matrix elements in the symmetrized basis}. Several methods can be used for their computation. The simplest is
to use the general formalism, that is to perform the change of basis associated with the symmetrization process of states and
operators in equation (\ref{eqb.13}). For the general operators we obtain
\begin{eqnarray}\label{eqb.32}
\lefteqn{\langle\langle \lbrack N_b\,0\rbrack J_b m'' A_1 |{}^{[m_4\,-m_4]}\mathcal{B}^{(j_b)}_{r A_1} |\lbrack
N_b\,0\rbrack J_b m' A_1 \rangle\rangle = } \nonumber \\
&&
(2J_b+1)^{-\frac{1}{2}} F\begin{array}{ccc} r A_1 & m' A_1 & ([N_b\,0]J_b)\ast \\
([m'_{4}\,-m'_{4}]j_b & [N_b\,0]J_b ) & m'' A_1
\end{array} \nonumber \\
&&\times \left([N_b\,0]J_b||{}^{[m_4\,-m_4]}\mathcal{B}^{(j_b)}{}||[N_b\,0]J_b\right),
\end{eqnarray}
where the \textit{rme} are those of equation (\ref{eqb.14b}); $r$ stands for $r_{\pm}=|m|\varepsilon$ or $0$ according to the
case and the symmetry adapted CG coefficients are given by:
\begin{eqnarray}\label{eqb.33}
\lefteqn{ F\begin{array}{ccc} r\,A_1 & m' A_1 & ([N_b\,0]J_b) \\
([m'_{4}\,-m'_{4}]j_b & [N_b\,0]J_b ) & m'' A_1
\end{array} =}\nonumber \\
&&\sum\limits_{m,m',m''} {~}^{[m'_{4}\,-m'_{4}]} G_{r\,A_1}^{m\,\ast} {~}^{[N_b\,0]} G_{m' A_1}^{m'\,\ast} {~}^{[N_b\,0]}
G_{m'' A_1}^{m''}  \nonumber \\
&& \times F\begin{array}{ccc} m & m' & ([N_b\,0]J_b) \\
([m'_{4} -m'_{4}]j_b & [N_b\,0]J_b ) & m'' \end{array} \nonumber \\
&&= \sum\limits_{m}{~}^{[m'_{4}\,-m'_{4}]} G_{p\,A_1}^{m\,\ast} \nonumber \\
&& \times F\begin{array}{ccc} m & m' & ([N_b\,0]J_b) \\
([m'_{4} -m'_{4}]j_b & [N_b\,0]J_b ) & m'' \end{array},
\end{eqnarray}
where the last equality follows from (\ref{eqb.22}). The analytical expressions for these coefficients together with those of the
various matrix elements are given in Appendix \ref{appa}.
\subsection{Stretching modes $\nu_1, \nu_3$}\label{stretch}
\subsubsection{Standard states and operators}\label{stretch1}
For the $su(2)_S$ Schwinger's realization we take:
\begin{equation}\label{eqs.1}
\begin{array}{l}
{}^{(s)}J_+ = b_1^+b_2 ~~,~~
{}^{(s)}J_- = b_2^+b_1 \\
{}^{(s)}J_z=  \frac{1}{2} (N_1-N_2) =\frac{1}{2} (b_1^+b_1-b_2^+b_2)
\end{array}
\end{equation}
where the indices $i=1,2$ are linked to the two bonds.\\
With regard to the notations of section \ref{sec22} we make the substitutions $N \rightarrow N_s$, $j \rightarrow j_s$ for the
states and $T\rightarrow S$ for the operators. We have then with (\ref{eq2.2.8}) ($j_s=n_s/2$):
\begin{equation}\label{eqs.1}
| [N_s\, \dot{0}]\lbrack n_s\,0\rbrack j_s\, m \rangle\rangle= \mbox{{\Large S}} {\renewcommand{\arraystretch}{1}
{\renewcommand{\arraycolsep}{-0.05cm} \left\lbrack
\begin{array}{c}
n_s \ 0 \ -n_s \\ \lbrack n_s \ 0 \rbrack \ (j_s) \\
m
\end{array}
\right\rbrack}}\ | [N_s\,\dot{0}] \lbrack 0\,0\rbrack 0 0 \rangle\rangle ,
\end{equation}
with
\[
| [N_s\, \dot{0}]\lbrack 0\,0\rbrack 0 0 \rangle\rangle \equiv |0,0,N_s \rangle =( N_s!)^{-1/2}b_3^{+N_s}|0,0,0\rangle .
\]
We may also express these states in various forms replacing the $S$ operator in (\ref{eqs.1}):
\begin{eqnarray}\label{eqsk.1}
\lefteqn{| [N_s\, \dot{0}]\lbrack n_s\,0\rbrack j_s\, m \rangle\rangle= (n_s!)^{1/2} \mathcal G(n_s,n_s,0)\
^{[n_s\,0]}S^{(j_s)}_{~m}|0,0\rangle |N_s\rangle }\nonumber \\
&&= (n_s!)^{1/2}\ \mathcal G(n_s,n_s,0)\ |[n_s\,0] j_s\,m \rangle \rangle |N_s\rangle \nonumber \\
&&=(-1)^{\frac{n_s}{2}+m}\, i^{n_s}\, |j_s-m,j_s+m,N_s-n_s\rangle .
\end{eqnarray}
We thus have the correspondence:
\begin{equation}\label{eqsk.2}
\frac{n_s}{2}-m=n_1~~,~~\frac{n_s}{2}+m=n_2~~,~~ N_s-n_s=n_3 .
\end{equation}
Arbitrary vibrational operators in the algebraic standard $u(3)\supset u(2)\supset su(2)\supset so(2)$ chain are obtained from
relations (\ref{eq2.2.4}, \ref{eq2.2.5}) and the results for the covariant operators $^{[m_1\,-m_2]}T^{(j)}_{~m}$. All their
matrix elements are given by equations (\ref{eq2.2.9}, \ref{eq2.2.10}). In particular, operators which are ``diagonal in $n_s$''
are characterized by $m_1=m_2$.
\subsubsection{Symmetrization in $C_{2v}$}\label{stretch2}
The computations have been made with the conventions given in Table \ref{tab1}.
\begin{table}[h]
\caption[$C_{2v}$ character table]{$C_{2v}$ character table} \label{tab1} \small{
$$
\begin{array}{c|cccc}
\hline
C_{2v}&E&C_{2}(Oz)&\sigma_{v}(xz)&\sigma_{v}(yz)\\
\hline
A_1&1&~1&~1&~1\\
  &  &  &  &    \\
A_2&1&~1&-1&-1\\
  &  &  &  &    \\
B_1&1&-1&~1&-1\\
  &  &  &  &    \\
B_2&1&-1&-1&~1\\
\hline
\end{array}
$$
} \normalsize
\end{table}
The indices $i=1,2$ being associated with the two bonds we have:
\begin{equation}\label{eqskx}
P_{C_2}\,b_1\,P_{C_2}^{-1}=b_2~;~P_{\sigma}\,b_1\,P_{\sigma}^{-1}=b_2~;~ P_{\sigma'}\,b_i\,P_{\sigma'}^{-1}=b_i ,
\end{equation}
where we set $\sigma=\sigma_{v}(yz)$ and $\sigma'=\sigma_{v}(xz)$. Hence the set $(b_1,b_2)$ (or $(b_1^+,b_2^+)$) span a
representation $A_1+B_1$ of $C_{2v}$.\\
$\bullet$ \textit{Symmetrized operators}. The relations in (\ref{eqskx}) allow first to determine the transformation laws of the
standard tensors (\ref{eq2.1.8}, \ref{eq2.1.11a}, \ref{eq2.2.5}) ($j=(m_1+m_2)/2$):
\begin{equation}\label{eqs.8}
P_{R}\,^{[m_1\,-m_2]}S^{(j)}_{~m}\,P_{R}^{-1}= (-1)^{m_1}\,^{[m_1\,-m_2]}S^{(j)}_{-m} ,
\end{equation}
where $R=C_2$, $\sigma$; they are obviously invariant with respect to $\sigma'$. We may thus build the symmetry adapted tensors
($m > 0$)
\begin{equation}\label{eqs.10}
\begin{split}
^{[m_1\,-m_2]}S^{(j)}_{|m|A_1}=\frac{\theta(m_1,m_2,|m|,\Gamma)}{\sqrt{2}}\,(\,^{[m_1\,-m_2]}S^{(j)}_{~m}
\\+(-1)^{m_1+\varepsilon}\,^{[m_1\,-m_2]}S^{(j)}_{-m}),
\end{split}
\end{equation}
with $\varepsilon=0$ for $\Gamma=A_1$ and $\varepsilon=1$ for $\Gamma=B_1$. For $m=0$ we simply have
\begin{equation}\label{eqs.11}
^{[m_1\,-m_2]}S^{(j)}_{0 \Gamma}=\theta(m_1,m_2,0,\Gamma)\,^{[m_1\,-m_2]}S^{(j)}_{~0},
\end{equation}
with
\begin{flalign}\label{eqs.12}
\Gamma&=A_1\ \mbox{for $m_1$ and $m_2$ even} \nonumber \\
\Gamma&=B_1\ \mbox{for $m_1$ and $m_2$ odd}.
\end{flalign}
In equations (\ref{eqs.10}, \ref{eqs.11}) $\theta$ is a phase factor to be fixed next.\\
In order to obtain a correct description (in terms of allowed symmetries in $C_{2v}$) for the states associated with the
\textit{irrep} $\lbrack N\,\dot{0}\rbrack$ of $u(3)$, we must impose that $b_3^+$ (or $b_3$) belongs to the $A_1$ scalar
representation of $C_{2v}$. With these conventions, it appears that the $\mathcal{G}(z,m_1,m_2)$ term , given by (\ref{eq2.2.4}),
is invariant in $C_{2v}$. Consequently the general operators (\ref{eq2.2.5}) transform, under the action of the $C_{2v}$
generators, as the standard operators (relation (\ref{eqs.8})). This property, allows to symmetrize both type of operators with
the same orientation matrix and we set for the $u(3)$ symmetrized operators:
\begin{eqnarray}\label{eqs.24}
\lefteqn{\mbox{{\Large S}} {\renewcommand{\arraystretch}{1} {\renewcommand{\arraycolsep}{-0.05cm} \left\lbrack
\begin{array}{c}
z\ \ \ 0 \ \ \ -z \\ \lbrack m_1 \, -m_2 \rbrack \, (j) \\
|m|\Gamma
\end{array}
\right\rbrack}} = }\nonumber \\
&&\sum\limits_m \,^{[m_{1}\,-m_{2}]} G_{|m|\Gamma}^{~m}\, \mbox{{\Large S}} {\renewcommand{\arraystretch}{1}
{\renewcommand{\arraycolsep}{-0.05cm} \left\lbrack
\begin{array}{c}
z\ \ \ 0 \ \ \ -z \\ \lbrack m_1 \, -m_2 \rbrack \, (j) \\
m
\end{array}
\right\rbrack}},
\end{eqnarray}
where the sum is limited to the values $m=\pm |m|$ ($m\neq 0$).\\ The choices for the phase factors have been made so that the
operators (\ref{eqs.24}) satisfy:
\begin{equation}\label{eqs.24c}
\mbox{{\Large S}} {\renewcommand{\arraystretch}{1} {\renewcommand{\arraycolsep}{-0.05cm} \left\lbrack
\begin{array}{c}
z\ \ \ 0 \ \ \ -z \\ \lbrack m_1 \, -m_2 \rbrack \ (j) \\
|m|\Gamma
\end{array}
\right\rbrack}}^{\dag} =\mbox{{\Large S}} {\renewcommand{\arraystretch}{1} {\renewcommand{\arraycolsep}{-0.05cm} \left\lbrack
\begin{array}{c}
z\ \ \ 0 \ \ \ -z \\ \lbrack m_2 \, -m_1 \rbrack \ (j) \\
|m|\Gamma
\end{array}\right\rbrack}}.
\end{equation}
Under time reversal they are unchanged  when $m_1$ or $m_2$ are zero and multiplied by $(-1)^{j+|m|+\varepsilon}$ when
$m_1,m_2\neq 0$. This leads to
\begin{equation}\label{eqs.24a}
\begin{split}
{~}^{[m_{1}\,-m_{2}]} G_{|m|\Gamma}^{~m}&=\frac{1}{\sqrt{2}}\,\theta(m_1,m_2,|m|,\Gamma)~~ ,m>0\\
{~}^{[m_{1}\,-m_{2}]} G_{|m|\Gamma}^{-m}&=\frac{1}{\sqrt{2}}\,\theta(m_1,m_2,|m|,\Gamma)(-1)^{m_1+\varepsilon}\\
{~}^{[m_{1}\,-m_{2}]} G_{|0|\Gamma}^{~0}&=\theta(m_1,m_2,0,\Gamma) ~~ ,m=0
\end{split}
\end{equation}
with
\begin{equation}\label{eqs.24b}
\begin{split}
\theta(m_1,m_2,|m|,\Gamma)&=i^{m_1}\,i^{j+|m|+\varepsilon}~~ m_1,m_2 \neq 0\\
\theta(m_1,0,|m|,\Gamma)&=i^{m_1}\\
\theta(0,m_2,|m|,\Gamma)&=i^{m_2}\,(-1)^{|m|+\varepsilon}.
\end{split}
\end{equation}
With the results in this section, keeping terms which are diagonal in $n_s$ only, we may write the effective stretching
Hamiltonian:
\begin{equation}\label{eqs.39}
\mathcal{H}_{S}= \displaystyle{\sum_{z=0}^{N_s}}\,\displaystyle{\sum_{m_1=0}^{z}}\,\displaystyle{\sum_{|m_s|}}^{*}
\,^{\{z\}}\tilde{t}_{|m_s|}^{(m_1)}\,\mbox{{\Large S}} {\renewcommand{\arraystretch}{1} {\renewcommand{\arraycolsep}{-0.05cm}
\left\lbrack
\begin{array}{c}
z\ \ \ 0 \ \ \ -z \\ \lbrack m_1\,  -m_1 \rbrack \ (j_s) \\
|m_s|\,A_1
\end{array}
\right\rbrack}},
\end{equation}
with $j_s=m_1$ and where the sum $\Sigma^{*}$ is over $|m_s|$ values such that $j_s+|m_s|=m_1+|m_s|$ be even.\\
$\bullet$ \textit{Symmetrized states}. These are obtained with equations (\ref{eqs.1}, \ref{eqsk.1}, \ref{eqsk.2}, \ref{eqs.24}):
\begin{equation}\label{eqs.ss}
| [N_s\, \dot{0}]\lbrack n_s\,0\rbrack j_s |m|\Gamma \rangle\rangle= \mbox{{\Large S}} {\renewcommand{\arraystretch}{1}
{\renewcommand{\arraycolsep}{-0.05cm} \left\lbrack
\begin{array}{c}
n_s \ 0 \ -n_s \\ \lbrack n_s \, 0 \rbrack \ (j_s) \\
|m|\Gamma
\end{array}
\right\rbrack}}\ | [N_s\,\dot{0}] \lbrack 0\,0\rbrack 0 0 \rangle\rangle ,
\end{equation}
with $^{[n_{s}\,0]} G$ obtained from (\ref{eqs.24a}, \ref{eqs.24b}). Explicitly we have for the various types of local states:\\
\textit{Local states $\{n_1,n_2\}$} $n_1=n_2$.\\ They are associated with $m=0$ (Eq. \ref{eqsk.2}) which implies $j_s$ integer or
(as $j_s=n_s/2$) $n_s$ even.
\begin{equation}\label{eqsk.1a}
| [N_s\, \dot{0}]\lbrack n_s\,0\rbrack j_s\, 0\,A_1 \rangle\rangle=
(-1)^{\frac{n_s}{2}}|\frac{n_s}{2},\frac{n_s}{2},N_s-n_s\rangle .
\end{equation}
\textit{Local states $\{n_1,n_2\}$ $n_1\neq n_2$}.\\ We have then $|m| \neq 0$ and still setting $\varepsilon =0$ for
$\Gamma=A_1$ and $\varepsilon =1$ for $\Gamma=B_1$:
\begin{eqnarray}\label{eqsk.1b}
\lefteqn{| [N_s\, \dot{0}]\lbrack n_s\,0\rbrack j_s\, |m|\Gamma \rangle\rangle= } \nonumber\\
&& \frac{(-1)^{\frac{n_s}{2}-|m|}}{\sqrt{2}} \{|\frac{n_s}{2}-m,\frac{n_s}{2}+m,N_s-n_s\rangle \nonumber
\\&&+(-1)^{\varepsilon}|\frac{n_s}{2}+m,\frac{n_s}{2}-m,N_s-n_s\rangle \}.
\end{eqnarray}
With our phase convention they are all invariant under time reversal.\\
$\bullet$ \textit{Matrix elements in the symmetrized basis}. They are obtained with a method similar to that used for the bending
mode. From equation (\ref{eq2.2.9}) the transformation to symmetrized states and operators gives
\begin{eqnarray}\label{eqs.28}
\lefteqn{ \langle\langle \Psi'' |{} \mbox{{\Large S}} {\renewcommand{\arraystretch}{1} {\renewcommand{\arraycolsep}{-0.05cm}
\left\lbrack
\begin{array}{c}
z\ \ \ 0 \ \ \ -z \\ \lbrack m_1 \, -m_2 \rbrack \ (j) \\
|m|\Gamma
\end{array}
\right\rbrack}}{}|\Psi' \rangle \rangle =(2j''+1)^{-1/2} }  \nonumber\\
& \times F\begin{array}{ccc} |m| \Gamma& |m'| \Gamma' & ([n_s''\,0]j_s'')\ast \\
([m_{1}\,-m_{2}]j & [n_s'\,0]j_s' ) & |m''| \Gamma''
\end{array} \, \nonumber\\
&\times\!\! \left([N_s\dot{0}][n_s''\,0]j_s''||{}\mbox{{\Large S}} {\renewcommand{\arraystretch}{1}
{\renewcommand{\arraycolsep}{-0.05cm} \left\lbrack
\begin{array}{c}
z\ \ \ 0 \ \ \ -z \\ \lbrack m_1 \, -m_2 \rbrack \ (j)
\end{array}
\right\rbrack}}||[N_s\,\dot{0}][n_s'\,0]j_s'\right)
\end{eqnarray}
where we set $|\Psi \rangle \rangle =|[N_s\,\dot{0}]\,[n_s\,0]j_s|m|\Gamma \rangle\rangle$ and the \textit{rme} are given by
(\ref{eq2.2.10}) with the appropriate label substitutions. The symmetry adapted CG coefficients are obtained with
\begin{eqnarray}\label{eqs.29}
\lefteqn{ F\begin{array}{ccc} |m| \Gamma& |m'| \Gamma' & ([n_s''\,0]j_s'') \\
([m_{1}\,-m_{2}]j & [n_s'\,0]j_s' ) & |m''| \Gamma''
\end{array} =} \nonumber \\
 & & \times \sum\limits_{m,m',m''} {~}^{[m_{1}\,-m_{2}]} G_{|m|\Gamma}^{m\,\ast} {~}^{[n'_s\,0]} G_{|m'|\Gamma'}^{m'\,\ast}
{~}^{[n''_s\,0]}
G_{|m''|\Gamma''}^{m''}  \nonumber \\ &&\times\, F\begin{array}{ccc} m & m' & ([n''_s\,0]j''_s) \\
([m_{1} -m_{2}]j & [n'_s\,0]j'_s ) & m'' \end{array},
\end{eqnarray}
with $j=(m_1+m_2)/2$, $j'_s=n'_s/2$ and $j''_s=n''_s/2$. Also it is important to note that these coefficients are \textit{a
priori} defined only for $n''_s=n'_s+m_1-m_2$. The matrices ${~}^{[m_{1}\,-m_{2}]} G$ and ${~}^{[n_{s}\,0]} G$ are given by
equations (\ref{eqs.24a}, \ref{eqs.24b}). The analytical expressions for these coefficients are given in Appendix \ref{appb}.
\subsection{Stretch-bend interactions}
Results in sections \ref{bend} et \ref{stretch} determine all operators adapted to the study of isolated bending and stretching
modes and which may appear in the Hamiltonian or transition moments.\\
Taking into account stretch-bend interactions introduces coupling operators which may be formally written
\[
\mathcal{O}_{sb}^{(\Gamma_{sb})}=[\mathcal{S}^{(\Gamma_{s})} \times \mathcal{B}^{(\Gamma_{b})}]^{(\Gamma_{sb})}.
\]
For our problem $\Gamma_{b}=A_1$ and the CG for the $C_{2v}$ group are trivial ; we thus simply have
\begin{equation}\label{eqsb.1}
\mathcal{O}_{sb}^{(\Gamma_{s})}=\mathcal{S}^{(\Gamma_{s})} \, \mathcal{B}^{(A_1)},
\end{equation}
with $\Gamma_{s}=A_1$ for Hamiltonian terms. The operators (\ref{eqsb.1}) may be written in various manners depending on the case
and also depending on what we mean to represent, for instance operators in the untransformed Hamiltonian or effective ones.
\subsubsection{Method in the case of a polyad structure}
The chosen dynamical algebra assumes $\omega_1 \approx \omega_3 (= \omega_s)$ and we have to take into account in the effective
Hamiltonian the resonance with the bending mode which determines the polyad structure. We assume
\[
\omega_1 \approx \omega_3 = \omega_s \approx k\, \omega_2=k \,\omega_b .
\]
Within our formalism the operator $N_1+N_2=\hat{n}_s$, with eigenvalue $n_s$, represents the ``number of quanta of stretching''
operator ; $N_4$ with eigenvalue $n_4=n_b$ the ``number of quanta of bending'' operator with
\begin{equation}\label{eqsb.2}
N_4\equiv \hat{n}_b= \frac{N_4+N_5}{2}+\frac{N_4-N_5}{2}=\frac{N_b}{2}+\,^{(b)}J_z .
\end{equation}
To a given $P$ polyad  we may associate the $\widehat{P}$ operator
\begin{equation}\label{eqsb.3}
\widehat{P}=\hat{n}_b+k\, \hat{n}_s =N_4+k\,(N_1+N_2),
\end{equation}
which may be expressed in terms of the  $ITO$ defined previously knowing that:
\[
N_4=\sqrt{2}(\, ^{[1\,-1]}\mathcal{B}^{(0)}_{0A_1}+\,^{[1\,-1]}B^{(1)}_{0A_1});
\]
\[
N_1+N_2= \frac{2}{3}\{ \,N_s- \sqrt{\frac{3N_s}{2}}\,\mbox{{\Large S}} {\renewcommand{\arraystretch}{1}
{\renewcommand{\arraycolsep}{-0.05cm} \left\lbrack
\begin{array}{c}
1\ \ \ 0 \ \ \ -1 \\ \lbrack 0\,  ~0 \rbrack \ (0) \\
0\, A_1
\end{array}
\right\rbrack}}\,\}.
\]
An $\mathcal{O}$ operator which conserves the $P$ quantum number associated with a given polyad must satisfy the condition
$[K,\mathcal{O}]=0$, be of species $A_1$ and invariant upon time reversal if it belongs to the Hamiltonian expansion.\\
To determine the possible $\mathcal{O}$ operators it appears that it is better to work first in the standard algebraic chain
\begin{equation*}\label{eqsb.4}
\begin{array}{cccccccc}
u(3)_S&\otimes&u(2)_B&\supset&u(2)_S&\otimes&u(2)_B&\supset\\
\,[N_s \dot{0}]&&[N_b0]&&[n_s 0]&&[N_b0]\\ \\
su(2)_S&\otimes&su(2)_B&\supset&so(2)_S&\otimes&so(2)_B\\
j_s=n_s/2&&j_b=N_b/2&&m_s&&m_b
\end{array}
\end{equation*}
where the indicated symmetries are those of the states. We thus start from the operator basis
\begin{equation}\label{eqsb.5}
\mathcal{S}\mathcal{B}=\mbox{{\Large S}} {\renewcommand{\arraystretch}{1} {\renewcommand{\arraycolsep}{-0.05cm} \left\lbrack
\begin{array}{c}
z\ \ \ 0 \ \ \ -z \\ \lbrack m_1\,  -m_2 \rbrack \ (j_s) \\
~m_s
\end{array}
\right\rbrack}}\ ^{[m_4\,-m_4]}\mathcal{B}^{(j_b)}_{~m_b},
\end{equation}
with $j_b=m'_4$ and $j_s=(m_1+m_2)/2$. A straightforward calculation gives
\[
[\widehat{P},\mathcal{S}\mathcal{B}]=[-m_b+k(m_1-m_2)]\mathcal{S}\mathcal{B},
\]
thus the condition $[\widehat{P},\mathcal{S}\mathcal{B}]=0$ is satisfied if
\begin{equation}\label{eqsb.6}
m_b=k\,(m_1-m_2),
\end{equation}
with $k=2$ for the considered $XY_2$ molecules. This condition being independent of $m_s$, it appears that, in order to determine
hermitian interaction operators having also a determined behavior upon time reversal, it is preferable to keep the standard form
for the bending operators and to take symmetrized operators for the stretching ones. We mainly have two cases:\\
$\bullet$ $m_b=0$ then $m_1=m_2$\\
In this case we can also take directly symmetrized bending operators, which gives the hermitian operators
\begin{equation}\label{eqsb.7}
\mathcal{O}_{sb}(1)=\mbox{{\Large S}} {\renewcommand{\arraystretch}{1} {\renewcommand{\arraycolsep}{-0.05cm} \left\lbrack
\begin{array}{c}
z\ \ \ 0 \ \ \ -z \\ \lbrack m_1\,  -m_1 \rbrack \ (j_s) \\
|m_s|\Gamma_s
\end{array}
\right\rbrack}}\ ^{[m_4\,-m_4]}\mathcal{B}^{(j_b)}_{0A_1},
\end{equation}
satisfying upon time reversal (see sections (\ref{bend2}, \ref{stretch2}))
\[
\mathcal{K}\mathcal{O}_{sb}(1)\mathcal{K}^{-1}=(-1)^{m_1+|m_s|+\varepsilon}\,\mathcal{O}_{sb}(1).
\]
$\bullet$ $m_b\neq 0$\\
>From the set (\ref{eqsb.5}) we may build the hermitian operators (equations (\ref{eq2.1.12}, \ref{eqs.24c}))
\begin{eqnarray} \label{eqsb.8}
\mathcal{O}_{sb}(2)=\frac{i^{\tau}}{\sqrt{2}}\left\{\mbox{{\Large S}} {\renewcommand{\arraystretch}{1}
{\renewcommand{\arraycolsep}{-0.05cm} \left\lbrack
\begin{array}{c}
z\ \ \ 0 \ \ \ -z \\ \lbrack m_1\,  -m_2 \rbrack \ (j_s) \\
|m_s|\Gamma_s
\end{array}
\right\rbrack}}\ ^{[m_4\,-m_4]}\mathcal{B}^{(j_b)}_{~m_b} \right. \nonumber \\
+\left.(-1)^{\tau} (-1)^{j_b+m_b}\,\mbox{{\Large S}} {\renewcommand{\arraystretch}{1} {\renewcommand{\arraycolsep}{-0.05cm}
\left\lbrack
\begin{array}{c}
z\ \ \ 0 \ \ \ -z \\ \lbrack m_2\,  -m_1 \rbrack \ (j_s) \\
|m_s|\Gamma_s
\end{array}
\right\rbrack}}\ ^{[m_4\,-m_4]}\mathcal{B}^{(j_b)}_{-m_b} \right\},
\end{eqnarray}
with $\tau=0,1$ and where $m_b>0$ and $m_1>m_2$ is assumed. With the properties established in sections (\ref{bend2},
\ref{stretch2})) one shows that
\begin{equation*}\label{eqsb.9}
\mathcal{K}\mathcal{O}_{sb}(2)\mathcal{K}^{-1}=\left\{ \begin{array}{ll} (-1)^{\tau}
(-1)^{j_b}(-1)^{j_s+|m_s|+\varepsilon}\mathcal{O}_{sb}(2)&(a)\\
(-1)^{\tau} (-1)^{j_b}\mathcal{O}_{sb}(2)&(b)
\end{array}\right.
\end{equation*}
where case $(a)$ (resp. $(b)$) is for $m_2\neq 0$ (resp. $m_2= 0$).
\subsubsection{First operators in the Hamiltonian for $k=2$}
Since the operators $\mathcal{B}$ only differ from the $B$ ones through a constant function within the \textit{irrep} $[N_b0]$,
we can make in equations (\ref{eqsb.7}, \ref{eqsb.8}) the substitutions
\[
^{[m_4\,-m_4]}\mathcal{B}^{(j_b)}_{~m_b} \rightarrow \,^{[m'_4\,-m'_4]}B^{(j_b)}_{~m_b}~~~~(j_b=m'_4),
\]
to define the terms (and the parameters) of the effective Hamiltonian. \\
$\bullet$ For $m_b = 0$ (and $m_4\neq 0$, $z\neq0$) the operators (\ref{eqsb.7}), with $\Gamma_s=A_1$ and
$(-1)^{m_1+|m_s|+\varepsilon}=1$ are products of operators belonging to $\mathcal{H}_B$ and $\mathcal{H}_S$.\\
$\bullet$ The cases $m_b \neq 0$ correspond to non trivial (that is non diagonal) interaction operators. With $k=2$ and as $m_1$
and $m_2$ are integers the even values of $m_b$ alone are allowed in (\ref{eqsb.8}), with a minimum value $m_{b\,min}=2$. As a
result $m'_{4\,min}=j_{b\,min}=2$ and $m_{4\,min}=2$. We have then $(m_1-m_2)_{min}=1$ from which we deduce $j_{s\,min}=1/2$ and
$|m_{s\,min}|=1/2$. We thus have the first allowed values for the labels in the operators (\ref{eqsb.8}):\\
\textit{i)} $m_1=1$, $m_2=0$, $j_s=1/2$, $|m_s|=|1/2|$ $\Rightarrow \tau=0$\\
\textit{ii)} $m_1=2$, $m_2=1$, $j_s=3/2$, $|m_s|=|1/2|$ $\Rightarrow \tau=0$\\
\textit{iii)} $m_1=2$, $m_2=1$, $j_s=3/2$, $|m_s|=|3/2|$ $\Rightarrow \tau=1$\\
We thus obtain:
\begin{equation}\label{eqsb.10}
\mathcal{O}_{sb}(2)_1=\frac{1}{\sqrt{2}}\left\{\mbox{{\Large S}} {\renewcommand{\arraystretch}{1}
{\renewcommand{\arraycolsep}{-0.05cm} \left\lbrack
\begin{array}{c}
z\ \ \ 0 \ \ \ -z \\ \lbrack 1\,  ~0 \rbrack \ (\frac{1}{2}) \\
|\frac{1}{2}|A_1
\end{array}
\right\rbrack}}\ ^{[2\,-2]}B^{(2)}_{~2} + h.c.\right\},
\end{equation}
with $z=1,2 \cdots$. The operators with $z>1$ are anharmonicity corrections to those obtained for $z=1$.\\We find next:
\begin{equation}\label{eqsb.12}
\mathcal{O}_{sb}(2)_2=\frac{1}{\sqrt{2}}\left\{\mbox{{\Large S}} {\renewcommand{\arraystretch}{1}
{\renewcommand{\arraycolsep}{-0.05cm} \left\lbrack
\begin{array}{c}
z\ \ \ 0 \ \ \ -z \\ \lbrack 2\,  -1 \rbrack \ (\frac{3}{2}) \\
|\frac{1}{2}|A_1
\end{array}
\right\rbrack}}\ ^{[2\,-2]}B^{(2)}_{~2} + h. c.\right\},
\end{equation}
\begin{equation}\label{eqsb.13}
\mathcal{O}_{sb}(2)_3=\frac{i}{\sqrt{2}}\left\{\mbox{{\Large S}} {\renewcommand{\arraystretch}{1}
{\renewcommand{\arraycolsep}{-0.05cm} \left\lbrack
\begin{array}{c}
z\ \ \ 0 \ \ \ -z \\ \lbrack 2\,  -1 \rbrack \ (\frac{3}{2}) \\
|\frac{3}{2}|A_1
\end{array}
\right\rbrack}}\ ^{[2\,-2]}B^{(2)}_{~2} - h.c.\right\},
\end{equation}
with $z=2,3 \cdots$.\\We also have other possible operators $^{[m_4\,-m_4]}B^{(j_b)}_{\pm 2}$ with $m_4 > 2$ in
(\ref{eqsb.10}-\ref{eqsb.13}) ; we restrict here to those which are of lowest degrees in creation and annihilation operators and
which will be used in the next section.
%********************************************************************
%*************Section 4************************************************
%********************************************************************
\section{Application to $D_2S$}\label{apd2s}
\subsection{Effective stretching Hamiltonian}
Up to the second order, that is for $z \leq 2$, the general expansion (\ref{eqs.39}) involves terms with the following values for
the labels
\[
\begin{array}{llll}
z=0&m_1=0&j_s=0&|m_s|=0 \\
z=1&m_1=0&j_s=0&|m_s|=0 \\
&m_1=1&j_s=1&|m_s|=1 \\
z=2&m_1=0&j_s=0&|m_s|=0 \\
&m_1=1&j_s=1&|m_s|=1 \\
&m_1=2&j_s=2&|m_s|=0 \\
&m_1=2&j_s=2&|m_s|=2
\end{array}
\]
With the results of sections \ref{sec221}, \ref{stretch2}, $\mathcal{H}_{S}$ can be written in terms of elementary boson
operators as
\begin{eqnarray}\label{eqs.41}
\lefteqn{\mathcal{H}_{S}=\,^{\{0\}}\tilde{t}_{0}^{(0)}+
\,^{\{1\}}\tilde{t}_{0}^{(0)}\,\,\sqrt{\frac{2}{3N_s}}\,[N_3-\frac{1}{2}(N_1+N_2)] }\nonumber \\
&&+\,^{\{1\}}\tilde{t}_{1}^{(1)}\,\frac{1}{\sqrt{2N_s}}\,(b^{+}_1b_2+b^{+}_2b_1) \nonumber \\
&&+\,^{\{2\}}\tilde{t}_{0}^{(0)}\,\,\frac{\alpha_{2,N_s}}{\sqrt{30}}\, \lbrace 3N_3(N_3-1)-6(N_1+N_2)N_3 \nonumber \\
&&+(N_1+N_2)(N_1+N_2-1)\rbrace \nonumber \\
&&+\,^{\{2\}}\tilde{t}_{1}^{(1)}\, \frac{\alpha_{2,N_s}}{\sqrt{10}} \ \lbrace
4N_3-(N_1+N_2-1)\}(b^{+}_1b_2+b^{+}_2b_1)\nonumber \\
&&+ \,^{\{2\}}\tilde{t}_{0}^{(2)}\,
\frac{\alpha_{2,N_s}}{\sqrt{6}}\,(b^{+2}_2b^{2}_2+b^{+2}_1b^{2}_1-4b^{+}_1b^{+}_2b_1b_2) \nonumber \\
&&+\,^{\{2\}}\tilde{t}_{2}^{(2)}\,\frac{\alpha_{2,N_s}}{\sqrt{2}}\,(b^{+2}_1b^{2}_2+\,b^{+2}_2b^{2}_1),
\end{eqnarray}
with $\alpha_{2, N_S}=[2N_S(N_S-1)]^{-\frac{1}{2}}$.\\
For the fitting procedure, it can be rewritten to the more convenient form:
\begin{eqnarray}
\lefteqn{\mathcal{H}_S=\alpha_0 (N_1+N_2) + \alpha_1 (N_1^2+N_2^2)+\alpha_2 N_1 N_2+ \alpha_3 Y^{(A_1)} }\nonumber \\
&&+\alpha_4 (N_1+N_2) Y^{(A_1)} + \alpha_5 [\,Y^{(A_1)}\times Y^{(A_1)}]^{(A_1)},
\end{eqnarray}
with $Y^{(A_1)}=b^{+}_{1}b_2+b^{+}_{2}b_1$ and where we removed the part of $H_S$ which depends upon the operator $N_s$ only
 since the latter takes a constant value within the \textit{irrep} $[N\, \dot{0}]$ of $u(3)_S$. Also we set:
\begin{eqnarray}
\lefteqn{\alpha_0=-{}^{\{1\}}\tilde{t}_{\,0}^{\,(0)}\sqrt{\frac{3}{2N_S}} +{}^{\{2\}}\tilde{t}_{\,0}^{\,(0)}(1-6
N_S)\sqrt{\frac{2}{15}}\,\alpha_{2, N_S} }\nonumber \\&& -{}^{\{2\}}\tilde{t}_{\,0}^{\,(2)}\frac{\alpha_{2, N_S}}{\sqrt{6}}
-{}^{\{2\}}\tilde{t}_{\,2}^{\,(2)}\frac{\alpha_{2, N_S}}{\sqrt{2}}, \nonumber \\
&&\alpha_1=\alpha_{2, N_S}[{}^{\{2\}}\tilde{t}_{\,0}^{\,(0)}\sqrt{\frac{10}{3}}+{}^{\{2\}}\tilde{t}_{\,0}^{\,(2)}\frac{
1}{\sqrt{6}}], \nonumber \\
&&\alpha_2=\alpha_{2, N_S}\displaystyle{\sqrt{\frac{2}{3}}}[{}^{\{2\}}\tilde{t}_{\,0}^{\,(0)}2 \sqrt{5}
-{}^{\{2\}}\tilde{t}_{\,0}^{\,(2)}2 -{}^{\{2\}}\tilde{t}_{\,2}^{\,(2)}\sqrt{3}] ,\nonumber \\
&&\alpha_3={}^{\{1\}}\tilde{t}_{\,1}^{\,(1)}\frac{1}{\sqrt{2N_S}}+{}^{\{2\}}\tilde{t}_{\,1}^{\,(1)}\alpha_{2,
N_S}\frac{1}{\sqrt{10}}(4N_s+1), \nonumber \\
&&\alpha_4=-{}^{\{2\}}\tilde{t}_{\,1}^{\,(1)}\alpha_{2, N_S}\sqrt{\frac{5}{2}} ~~,~~
\alpha_5={}^{\{2\}}\tilde{t}_{\,2}^{\,(2)}\alpha_{2, N_S}\frac{1}{\sqrt{2}}. \nonumber
\end{eqnarray}
\subsection{Effective bending Hamiltonian}
The general expansion (\ref{eqb.39}) may also be written, using the results in section (\ref{bend2}):
\begin{equation*}\label{eqb.40}
\mathcal{H}_B=\!\sum_{m_4=0}^{N_b}\!\sum_{j_b=0}^{m_4}\,^{\{m_4\}}t_{b}^{(j_b)}\,{}^{\lbrace m_4,m_4\rbrace}\mbox{{\Large
g}}_{j_b}(N_4\!+\!N_5)  \, ^{[m'_4\,-m'_4]}B^{(j_b)}_{0 A_1},
\end{equation*}
with $j_b=m'_4$ and $^{\lbrace m_4,m_4\rbrace}\mbox{{\Large g}}_{j_b}$ defined in (\ref{eqb.11}). As a given $j_b$ (or $m'_4$)
value appears for all $m_4\geq j_b$ values and since $N_4+N_5$ takes a constant $N_b$ value within the \textit{irrep} $[N_b\,0]$
of $u(2)_B$ we may set
\begin{equation}\label{eqb.42}
\mathcal{H}_B=\displaystyle{\sum_{j_b=0}^{N_b}}\,\tilde{t}_{b}^{(j_b)}\,^{[m'_4\,-m'_4]}B^{(j_b)}_{0 A_1},
\end{equation}
with the effective parameters:
\begin{equation}\label{eqb.43}
\tilde{t}_{b}^{(j_b)}=\sum_{m_4=j_b}^{N_b}{}^{\{m_4\}}t_{b}^{(j_b)}\,{}^{\lbrace m_4,m_4\rbrace}\mbox{{\Large g}}_{j_b}(N_b).
\end{equation}
For instance up to second order we have:
\begin{eqnarray}\label{eqb.45}
\lefteqn{\mathcal{H}_B^{(2)}=\,\tilde{t}_{b}^{(0)}+\tilde{t}_{b}^{(1)}\,^{[1\,-1]}B^{(1)}_{0 A_1}+\tilde{t}_{b}^{(2)}\,^{[2\,-2]}B^{(2)}_{0 A_1}, } \\
&&=\,\tilde{t}_{b}^{(0)}+\frac{\tilde{t}_{b}^{(1)}}{\sqrt{2}}(N_4-N_5) \nonumber
\\&&+\frac{\tilde{t}_{b}^{(2)}}{2\sqrt{6}}\,[{}b^{+2}_4b^{2}_4+{}b^{+2}_5b^{2}_5-4b^{+}_4b^{+}_5b_4b_5],\nonumber
\nonumber \\
&&=\beta_0 N_4 + \beta_1 N_4^2
\end{eqnarray}
where the last form has a clearer physical meaning, with parameters $\beta_0$, $\beta_1$ given by:
\[
\beta_0=\sqrt{2}\,\tilde{t}_{\,b}^{\,(1)}-\sqrt{\frac{3}{2}}N_b \,\tilde{t}_{\,b}^{\,(2)}~~,~~ \beta_1=\sqrt{\frac{3}{2}}
\,\tilde{t}_{\,b}^{\,(2)}.
\]
\subsection{Effective stretch-bend interaction operators}
As noted before these interaction operators may be divided in two groups. In the first one we have products of stretching and
bending operators diagonal with respect to $n_s$ and $n_b$. In the second one are those obtained from the properties of the
$u(3)_S$ and $u(2)_B$ dynamical algebras and which take into account the approximate resonance between the stretching an bending
modes. Keeping only operators of lowest degree the stretch-bend effective Hamiltonian can be expressed as (equations
(\ref{eqsb.7}), (\ref{eqsb.10})):
\begin{equation}
 \mathcal{H}_{SB}^{I}=\gamma_0 (N_1+N_2) N_4 + \gamma_1 N_4\, Y^{(A_1)},
\end{equation}
\begin{equation}
 \mathcal{H}_{SB}^{II}=\gamma_2\,(b^{+}_{2}b^{}_{3}b^{+2}_{5}b^{2}_{4}+b^{+}_{1}b^{}_{3}b^{+2}_{5}b^{2}_{4}+
 h.c.)=\gamma_2\,O_{SB}^{II}.
\end{equation}
It is worth to analyze more deeply this last operator which indeed traduces the resonance $\nu_{1}(A_1) \simeq \nu_{3}(B_1)
\simeq  2\nu_{2}(A_1)$. Thus the Hamiltonian matrix, already divided into two $A_1$ and $B_1$ blocks , is subdivided into
sub-blocks characterized
by the polyad number $P=2(n_1+n_2)+n_b$. Within each $P$-block, $H_{SB}^{II}$ connects states which are not diagonal neither in $n_s$ nor in $n_b$.\\
$\mathcal{O}_{SB}^{II}$ can also be written $(b^{+}_{1}+ b^{+}_{2})b^{2}_{4}[b^{+2}_{5}b_3] +h.c.$. In this form it is clear that
the dependence of its matrix elements upon the quantum numbers $n_1$, $n_2$ and $n_b$ are similar to that of a usual Fermi
interaction operator. However contrarily to the later, which leads to convergence problems for high values of the quantum
numbers, the other factor the matrix elements of which behave roughly as $(N_s-n_s)^{1/2}(N_b-n_b)$, has a damping effect
as $n_s$ and (or) $n_b$ increase ($N_s$ and $N_b$ fixed).\\
We already defined a similar operator adapted to $XY_3$ pyramidal molecules in \cite{PLU2005,SAN2008} and already proved that
this operator does not require the knowledge of $N_s$ and $N_b$ for low values of the quantum numbers $n_1$, $n_2$ and $n_3$ as
it physically must be near the minimum of the potential function.
\subsection{Numerical application}
To illustrate the efficiency of our formalism, we apply it to the deuterate hydrogen sulfide. The hydrogen sulfide molecule and
its isotopic species are of interest for terrestrial atmospheric pollutant measurements. They are involved in the study of planet
atmospheres and appear in the analysis of interstellar medium. Also, hydrogen sulfide is a good candidate to apply local mode
models. We restrict here to the $D_2S$ molecule but will present a comparative analysis with other $XY_2$ molecules in a next paper.\\
Renaming, for simplicity, the parameters $a_i$ (i=0, ...,7), the best fitted Hamiltonian is defined as follows:
\begin{flalign}\label{GOOD_HAMILTONIAN}
H&=a_0 (N_1+N_2) + a_1 (N_1^2+N_2^2)+a_2 N_1 N_2 + a_3 Y^{(A_1)} \nonumber \\
 &+a_4 N_4 + a_5 N_4^2+a_6\, O_{SB}^{II}+a_7
(N_1+N_2)N_4.
\end{flalign}
However, taking into account the $N_s$ and $N_b$ quantum numbers, there are 10  parameters to determine in our model since the
value of these numbers appears explicitly in the matrix elements of $O_{SB}^{II}$. To obtain the optimum values of these quantum
numbers, we simply operate as follows. An initial set for the parameters $a_i^{(0)}$ ($i=0,1,4,5$) is obtained from the
experimental values of the lowest vibrational bands. This allows to have approximate values for the stretching levels when $n_s$
quanta are localized on one bond and for the bending levels:
\[
E_s^{(0)}(n_s)=a_0^{(0)}n_s+a_1^{(0)}n_s^2~,~E_b^{(0)}(n_b)=a_4^{(0)}n_b+a_5^{(0)}n_b^2.
\]
As for any anharmonic potential (Morse or modified P\"{o}sch-Teller for instance) the $n_{max}$ value of the vibrational quantum
number is given by the extremum of the $E_s^{(0)}(n_s)$ and $E_b^{(0)}(n_b)$  curves. With
\[
\begin{array}{ll}
a_0^{(0)}=1927.5855\ \mbox{cm}^{-1},&a_1^{(0)}=-24.4279 \ \mbox{cm}^{-1},\\
a_4^{(0)}=858.2604 \ \mbox{cm}^{-1},&a_5^{(0)}=-2.8564 \ \mbox{cm}^{-1},
\end{array}
\]
we obtained in this way:
\[
n_{s\,max}=-\frac{a_0^{(0)}}{2a_1^{(0)}}=39.45~,~n_{b\,max}=-\frac{a_4^{(0)}}{2a_5^{(0)}}=150.23,
\]
hence we could take $(N_s^{(0)},N_b^{(0)})=(39,150)$. However this high value for $N_b^{(0)}$ is not reasonable for the following
reasons. As it is commonly accepted $N_s$ may be interpreted as the number of excitation quanta which, when concentrated on one
bond, may dissociate the molecule. Within a polyad $P$ to which a $|N_s,0,0\rangle$ stretching state belongs, we also have states
$|0,0,N_s\rangle|2N_s,N_b-2N_s\rangle$ which implies that $N_b\geq 2N_s$, but on the other hand if we took $N_b^{(0)}$ much
greater than $2N_s^{(0)}$, we would have many dissociating states within polyads with $P\geq 2N_s^{(0)}$. It would be quite
unrealistic to pretend that our model is capable to reproduce isolated stable states within the continuum. Another method to
obtain a reasonable value for $N_s^{(0)}$ is to use for $E_s^{(0)}(n_s)$ the known experimental value of the dissociation energy
\cite{PEE2002}:
\[
a_0^{(0)}n_s+a_1^{(0)}n_s^2 \simeq 32050 \pm 50 \ cm^{-1},
\]
which leads to $N_s^{(0)'}=24$. Various fits were performed starting thus with $(N_s^{(0)},N_b^{(0)})=(39,78)$ while the other
$a_i$ ($i=0, \cdots ,7$) parameters  were determined through a usual non-linear least square fit method. We noticed that, except
for the $a_6$ parameter associated with $O_{SB}^{II}$ the matrix elements of which depend strongly upon the $N_s$ and $N_b$
values, other parameters remained almost unchanged (less than some percent of relative variation) while $N_s$ and $N_b$
decreased. One of the indicator of the convergence of the fitting process was the minimization of the standard deviation
\[\sigma(d,p)=\sqrt{\displaystyle{\frac{1}{d-p}}\sum_{i=1}^d \left\lbrack E_i^{(cal)}-E_i^{(obs)} \right\rbrack ^2}, \]
where $d$ and $p$ are respectively the number of experimental data and the number of parameters included in the fit.\\
It soon appeared that this indicator was rather insensitive to the $N_b$ value in a rather large range. A similar effect was
already noticed in previous studies of pyramidal molecules \cite{SAN2008}. On the other hand using the initial
$(N_s^{(0)},N_b^{(0)})=(24,48)$ values improved drastically the convergence to the minimum value $\sigma(22,8)=0.514\mbox {
cm}^{-1}$, thus reducing noticeably the computational time.
\begin{table}[h!]
\caption[Parameters (in $\mbox{cm}^{-1}$) fitted with 22 experimental data.]{Parameters (in $\mbox{ cm}^{-1}$) fitted with 22
experimental data.} \label{good parameters}
\small{
$$
\begin{array}{c r r}
\hline
a_{0}      & = &       1927.908(290)     \\
 &  &            \\
a_{1}      & = &        -24.665(103)     \\
 &  &            \\
a_{2}      & = &         -0.845(288)    \\
 &  &            \\
a_{3}      & = &         -6.428(140)     \\
 &  &            \\
a_{4}      & = &        858.821(447)      \\
 &  &            \\
a_{5}      & = &          -3.103(141)      \\
 &  &            \\
a_{6}      & = &          0.005(001)      \\
 &  &            \\
a_{7}      & = &          -10.488(150)        \\
\hline
\end{array}
$$
} \normalsize
\end{table}
The second indicator is, of course, the parameters stability at the end of the fitting procedure; the last variation $\Delta a_i$
of the parameters fulfilled the condition $ |\Delta a_i/a_i|<10^{-7}$, $(0\leq i \leq 7)$. The retained Hamiltonian
({\ref{GOOD_HAMILTONIAN}}) led to the set of parameters given in Table \ref{good parameters}
(these are given with values in parentheses which are 1$\sigma$ statistical confidence intervals in units of the last digits).\\
Finally we compared the experimental dissociation energy with the value calculated from our model. This was done in two ways.
First, removing all off-diagonal terms in the Hamiltonian (\ref{GOOD_HAMILTONIAN}), the energy of the $|24, 0, 0\, A_1 \rangle$
(or $|24, 0, 0\, B_1 \rangle$) pure stretching state is computed with the $a_0$ and $a_1$ parameters of Table \ref{good
parameters} which leads to the dissociation energy $E_D=32063\mbox{ cm}^{-1}$, or more precisely if we take into account the
parameter uncertainties $31996\mbox{ cm}^{-1}<E_D<32129\mbox{ cm}^{-1}$. The second way to calculate $E_D$ is to diagonalize the
Hamiltonian (\ref{GOOD_HAMILTONIAN}) within the $P=48$ polyad with the parameters $a_i$ (i=0, ...,7) of Table \ref{good
parameters}. The value obtained for the $|24, 0, 0\, A_1\rangle$, $|24, 0, 0\,B_1\rangle$ levels is $E'_D=32062 \mbox{ cm}^{-1}$
which is $1 \mbox{ cm}^{-1}$ close to the previously calculated $E_D$ value, which confirms the validity of our previous
assumption to keep only diagonal operators to evaluate this dissociation energy. Taking again into account the uncertainty on the
parameters, we found that the diagonalization of the $P=48$ Hamiltonian matrix with the two sets:
\begin{flalign}
\{a_i(min)\} = \{a_0^{-}, a_1^{-}, a_2^{-}, a_3^{-}, a_4^{-}, a_5^{-}, a_6^{+}, a_7^{-} \} \nonumber \\
\{a_i(max)\} = \{a_0^{+}, a_1^{+}, a_2^{+}, a_3^{+}, a_4^{+}, a_5^{+}, a_6^{-}, a_7^{+} \}, \nonumber
\end{flalign}
where $a_i^{-}$ (resp. $a_i^{+}$) stands for the lowest (resp. highest) value of the $a_i$ parameter according to the 1$\sigma$
statistical confidence interval leads to
\[
E'_{D_{min}}= 31996\ \mbox{cm}^{-1} < E'_D < E'_{D_{max}}= 32128 \ \mbox{cm}^{-1}.
\]
Both methods lead to similar values also consistent with the experimental results.\\
Experimental and calculated energies of $D_2S$ vibrational levels are given in Table \ref{good_table}. The first column gives the
normal notation ($\nu_1$  $\nu_2$ $\nu_3$) of the level. The second one indicates the polyad $P$ number. The third  column
indicates the levels in local notation with explicit symmetric (+) and antisymmetric (-) labels. The usual local mode notation
has been adapted to our notation as follows $m n^{\pm}, v\approx  n_1 n_2^{\pm}, n_4\equiv n_b$. Column 4 gives the eigenvalues,
whereas column 5 indicates the observed energy levels. Column 6 gives the difference Observed-Calculated energy. The last column
shows that the eigenkets are close to the initial basis given in columns 1 and 3. We mention that all the experimental data used
in the present paper are reported in \cite{LIU2006}.
\begin{table*}[h]
\caption[Observed and calculated energies of $D_2S$ with 21 experimental data. ($1\leq P \leq 12$).]{Observed and calculated
energies of $D_2S$ with 22 experimental data. ($1\leq P \leq 12$).}
 \label{good_table}
\small{
\begin{center}
\begin{tabular}
%{cccc@{\hspace{2\tabcolsep}}c@{\hspace{2\tabcolsep}}r@{}l@{\hspace{2\tabcolsep}}r@{}l@{\hspace{2\tabcolsep}}r@{}l@{\hspace{2\tabcolsep}}c}\\
{ccccccr@{}lr@{}lr@{}lc}\\

 \hline \rule[-0.3cm]{0cm}{0.8cm} \vspace{-0.2cm}
&\multicolumn{3}{c}{Normal} &Polyad&    Local      & \multicolumn{2}{c}{E$_{cal}$} & \multicolumn{2}{c}{E$_{obs}$} & \multicolumn{2}{c}{E$_{obs.}$-E$_{cal.}$}&\%init.ket\\
\vspace{0.1cm}
&$\nu_1$ & $\nu_2$ &$\nu_3$ &$P$& $n_1n_2\pm,n_4$ & \multicolumn{2}{c}{$\mbox{cm}^{-1}$} & \multicolumn{2}{c}{$\mbox{cm}^{-1}$} & \multicolumn{2}{c}{$\mbox{cm}^{-1}$}  &$(Modulus)$ \\

\hline
&0& 1 & 0 &$P$=1 &$00^{+},$ 1 &  855.&71821&  855.&40416 &  -0.&31405 & 1.00000 \\
& &   &   &      &            &      &     &      &      &     &      &         \\

&0& 2 & 0 &$P$=2 &$00^{+},$ 2 & 1705.&19892&      &      &     &      & 0.99992 \\
&1& 0 & 0 &      &$10^{+},$ 0 & 1896.&84567& 1896.&43154 &  -0.&41413 & 0.99992 \\
&0& 0 & 1 &      &$10^{-},$ 0 & 1909.&67175& 1910.&18375 &   0.&51200 & 1.00000 \\
& &   &   &      &            &      &     &      &      &     &      &         \\

&0& 3 & 0 &$P$=3 &$00^{+},$ 3 & 2548.&44694& 2549.&07336 &   0.&62642 & 0.99977 \\
&1& 1 & 0 &      &$10^{+},$ 1 & 2742.&13327& 2742.&66570 &   0.&53243 & 0.99977 \\
&0& 1 & 1 &      &$10^{-},$ 1 & 2754.&90182& 2754.&45192 &  -0.&44990 & 1.00000 \\
& &   &   &      &            &      &     &      &      &     &      &         \\

&0& 4 & 0 &$P$=4 &$00^{+},$ 4 & 3385.&46874&      &      &     &      & 0.99958 \\
&1& 2 & 0 &      &$10^{+},$ 2 & 3583.&56418&      &      &     &      & 0.99941 \\
&0& 2 & 1 &      &$10^{-},$ 2 & 3593.&89050& 3593.&12888 &  -0.&76162 & 0.99989 \\
&2& 0 & 0 &      &$20^{+},$ 0 & 3754.&01   & 3753.&47    &  -0.&54    & 0.97041 \\
&1& 0 & 1 &      &$20^{-},$ 0 & 3757.&19161& 3757.&45948 &   0.&26787 & 0.99989 \\
&0& 0 & 2 &      &$11^{+},$ 0 & 3808.&85313& 3809.&15400 &   0.&30087 & 0.97049 \\
& &   &   &      &            &      &     &      &      &     &      &         \\

&0& 5 & 0 &$P$=5 &$00^{+},$ 5 & 4216.&26769&      &      &     &      & 0.99936 \\
&1& 3 & 0 &      &$10^{+},$ 3 & 4418.&02232& 4417.&95854 &  -0.&06378 & 0.99886\\
&0& 3 & 1 &      &$10^{-},$ 3 & 4426.&64338& 4426.&08293 &  -0.&56045 & 0.99970 \\
&2& 1 & 0 &      &$20^{+},$ 1 & 4588.&84137& 4589.&22600 &   0.&38463 & 0.97039 \\
&1& 1 & 1 &      &$20^{-},$ 1 & 4591.&99807& 4592.&18104 &   0.&18297 & 0.99970 \\
&0& 1 & 2 &      &$11^{+},$ 1 & 4643.&6204 & 4643.&4770  &  -0.&1434  & 0.97062 \\
& &   &   &      &            &      &     &      &      &     &      &         \\

&0& 6 & 0 &$P$=6 &$00{^+},$ 6 & 5040.&84790&      &      &     &      & 0.99911  \\
&1& 4 & 0 &      &$10{^+},$ 4 & 5246.&10774&      &      &     &      & 0.99813  \\
&0& 4 & 1 &      &$10{^-},$ 4 & 5253.&16560&      &      &     &      & 0.99944  \\
&2& 2 & 0 &      &$20{^+},$ 2 & 5417.&41884&      &      &     &      & 0.97002  \\
&1& 2 & 1 &      &$20^{-},$ 2 & 5420.&5503 & 5421.&3007  &   0.&7504  & 0.99919 \\
&0& 2 & 2 &      &$11{^+},$ 2 & 5472.&14111&      &      &     &      & 0.97056  \\
&3& 0 & 0 &      &$30^{+},$ 0 & 5560.&36   & 5560.&15    &  -0.&21    & 0.99146 \\
&2& 0 & 1 &      &$30^{-},$ 0 & 5560.&69   & 5560.&74    &   0.&05    & 0.99480 \\
&1& 0 & 2 &      &$21^{+},$ 0 & 5647.&40   & 5647.&13    &  -0.&27    & 0.99146 \\
&0& 0 & 3 &      &$21^{-},$ 0 & 5672.&69   & 5672.&89    &   0.&20    & 0.99500 \\
& &   &   &      &            &      &     &      &      &     &      &         \\
 \hline
\end{tabular}
\end{center}
} \normalsize
\end{table*}
\addtocounter{table}{-1}
\begin{table*}[tpb]
\caption{(cont.)}
 \small
\begin{center}
\begin{tabular}
%{cccc@{\hspace{2\tabcolsep}}c@{\hspace{2\tabcolsep}}r@{}l@{\hspace{2\tabcolsep}}r@{}l@{\hspace{2\tabcolsep}}r@{}l@{\hspace{2\tabcolsep}}c}\\
{ccccccr@{}lr@{}lr@{}lc}\\
\hline
&\multicolumn{3}{c}{Normal} &Polyad&    Local      & \multicolumn{2}{c}{E$_{cal}$} & \multicolumn{2}{c}{E$_{obs}$} & \multicolumn{2}{c}{E$_{obs.}$-E$_{cal.}$}&\%init.ket\\
\vspace{0.1cm}
&$\nu_1$ & $\nu_2$ &$\nu_3$ &$P$& $n_1n_2\pm,n_4$ & \multicolumn{2}{c}{$\mbox{cm}^{-1}$} & \multicolumn{2}{c}{$\mbox{cm}^{-1}$} & \multicolumn{2}{c}{$\mbox{cm}^{-1}$}  &$(Modulus)$ \\

\hline
&0& 7 & 0 &$P$=7 &$00{^+},$ 7 & 5859.&21285&      &      &     &      & 0.99885  \\
&1& 5 & 0 &      &$10{^+},$ 5 & 6067.&85466&      &      &     &      & 0.99728  \\
&0& 5 & 1 &      &$10{^-},$ 5 & 6073.&46187&      &      &     &      & 0.99912  \\
&2& 3 & 0 &      &$20{^+},$ 3 & 6239.&74543&      &      &     &      & 0.96935  \\
&1& 3 & 1 &      &$20{^-},$ 3 & 6242.&85529&      &      &     &      & 0.99844  \\
&0& 3 & 2 &      &$11{^+},$ 3 & 6294.&42020&      &      &     &      & 0.97032  \\
&3& 1 & 0 &      &$30^{+},$ 1 & 6384.&74   & 6384.&63    &  \ \ -0.&11    & 0.99114 \\
&2& 1 & 1 &      &$30^{-},$ 1 & 6385.&06   & 6384.&99    &  \ \ -0.&07    & 0.99443 \\
&1& 1 & 2 &      &$21{^+},$ 1 & 6471.&82444 &      &     &     &      & 0.99114\\
&0& 1 & 3 &      &$21{^-},$ 1 & 6496.&95890 &      &     &     &      & 0.99500\\
& &   &   &      &            &      &     &      &      &     &      &         \\
&0& 8 & 0 &$P$=8 &$00{^+},$ 8 & 6671.&36558 &      &      &    &       &0.99858\\
&1& 6 & 0 &      &$10{^+},$ 6 & 6883.&27937 &      &      &    &       &0.99633\\
&0& 6 & 1 &      &$10{^-},$ 6 & 6887.&53650 &      &      &    &       &0.99877\\
&2& 4 & 0 &      &$20{^+},$ 4 & 7055.&82918 &      &      &    &       &0.96846\\
&1& 4 & 1 &      &$20{^-},$ 4 & 7058.&91965 &      &      &    &       &0.99749\\
&0& 4 & 2 &      &$11{^+},$ 4 & 7110.&46700 &      &      &    &       &0.96998\\
&3& 2 & 0 &      &$30{^+},$ 2 & 7202.&89484 &      &      &    &       &0.99079\\
&2& 2 & 1 &      &$30{^-},$ 2 & 7203.&16108 &      &      &    &       &0.99347\\
&1& 2 & 2 &      &$21{^+},$ 2 & 7293.&19251 &      &      &    &       &0.99097\\
&0& 2 & 3 &      &$21{^-},$ 2 & 7314.&97003 &      &      &    &       &0.99102\\
&4& 0 & 0 &      &$40{^+},$ 0 & 7315.&94818 &      &      &    &       &0.99555\\
&3& 0 & 1 &      &$40{^-},$ 0 & 7315.&97536 &      &      &    &       &0.99155\\
&2& 0 & 2 &      &$31{^+},$ 0 & 7454.&90209 &      &      &    &       &0.92603\\
&1& 0 & 3 &      &$31{^-},$ 0 & 7463.&63256 &      &      &    &       &0.99601\\
&0& 0 & 4 &      &$22{^+},$ 0 & 7519.&73558 &      &      &    &       &0.93012\\
& &   &   &      &            &      &     &      &      &     &      &         \\
&0& 9 & 0 &$P$=9 &$00{^+},$ 9 & 7477.&30881 &      &      &    &       &0.99831\\
&1& 7 & 0 &      &$10{^+},$ 7 & 7692.&39210 &      &      &    &       &0.99533\\
&0& 7 & 1 &      &$10{^-},$ 7 & 7695.&39342 &      &      &    &       &0.99840\\
&2& 5 & 0 &      &$20{^+},$ 5 & 7865.&67458 &      &      &    &       &0.96740\\
&1& 5 & 1 &      &$20{^-},$ 5 & 7868.&74925 &      &      &    &       &0.99641\\
&0& 5 & 2 &      &$11{^+},$ 5 & 7920.&28460 &      &      &    &       &0.96955\\
&3& 3 & 0 &      &$30{^+},$ 3 & 8014.&77154 &      &      &    &       &0.98976\\
&2& 3 & 1 &      &$30{^-},$ 3 & 8014.&99941 &      &      &    &       &0.99205\\
&1& 3 & 2 &      &$21{^+},$ 3 & 8107.&35389 &      &      &    &       &0.99022\\
\hline
\end{tabular}
\end{center}
\end{table*}
\addtocounter{table}{-1}
\begin{table*}[tpb]
\caption{(cont.)}
 \small
\begin{center}
\begin{tabular}
%{cccc@{\hspace{2\tabcolsep}}c@{\hspace{2\tabcolsep}}r@{}l@{\hspace{2\tabcolsep}}r@{}l@{\hspace{2\tabcolsep}}r@{}l@{\hspace{2\tabcolsep}}c}\\
{ccccccr@{}lr@{}lr@{}lc}\\
\hline
&\multicolumn{3}{c}{Normal} &Polyad&    Local      & \multicolumn{2}{c}{E$_{cal}$} & \multicolumn{2}{c}{E$_{obs}$} & \multicolumn{2}{c}{E$_{obs.}$-E$_{cal.}$}&\%init.ket\\
\vspace{0.1cm}
&$\nu_1$ & $\nu_2$ &$\nu_3$ &$P$& $n_1n_2\pm,n_4$ & \multicolumn{2}{c}{$\mbox{cm}^{-1}$} & \multicolumn{2}{c}{$\mbox{cm}^{-1}$} & \multicolumn{2}{c}{$\mbox{cm}^{-1}$}  &$(Modulus)$ \\

\hline

&0& 3 & 3 &      &$21{^-},$ 3 & 8126.&74881 &      &      &    &       &0.99357\\
&4& 1 & 0 &      &$40{^+},$ 1 & 8129.&89865 &      &      &    &       &0.99477\\
&3& 1 & 1 &      &$40{^-},$ 1 & 8129.&92376 &      &      &    &       &0.99373\\
&2& 1 & 2 &      &$31{^+},$ 1 & 8268.&83916 &      &      &    &       &0.92604\\
&1& 1 & 3 &      &$31{^-},$ 1 & 8277.&50278 &      &      &    &       &0.99571\\
&0& 1 & 4 &      &$22{^+},$ 1 & 8333.&53587 &      &      &    &       &0.93038\\
& &   &   &      &            &      &     &      &      &     &      &         \\
&0&10 & 0 &$P$=10&$00{^+},$10 & 8301.&04495 &      &      &    &       &0.99805\\
&1& 8 & 0 &      &$10{^+},$ 8 & 8500.&20154 &      &      &    &       &0.99430\\
&0& 8 & 1 &      &$10{^-},$ 8 & 8506.&03621 &      &      &    &       &0.99801\\
&2& 6 & 0 &      &$20{^+},$ 6 & 8675.&26677 &      &      &    &       &0.96624\\
&1& 6 & 1 &      &$20{^-},$ 6 & 8677.&34945 &      &      &    &       &0.99523\\
&0& 6 & 2 &      &$11{^+},$ 6 & 8726.&87621 &      &      &    &       &0.96908\\
&3& 4 & 0 &      &$30{^+},$ 4 & 8822.&39006 &      &      &    &       &0.98833\\
&2& 4 & 1 &      &$30{^-},$ 4 & 8822.&58468 &      &      &    &       &0.99028\\
&1& 4 & 2 &      &$21{^+},$ 4 & 8910.&09790 &      &      &    &       &0.98909\\
&0& 4 & 3 &      &$21{^-},$ 4 & 8929.&29048 &      &      &    &       &0.99353\\
&4& 2 & 0 &      &$40{^+},$ 2 & 8938.&54404 &      &      &    &       &0.99280\\
&3& 2 & 1 &      &$40{^-},$ 2 & 8938.&56874 &      &      &    &       &0.99215\\
&2& 2 & 2 &      &$31{^+},$ 2 & 9020.&00917 &      &      &    &       &0.99635\\
&1& 2 & 3 &      &$31{^-},$ 2 & 9020.&01004 &      &      &    &       &0.99636\\
&0& 2 & 4 &      &$22{^+},$ 2 & 9077.&50764 &      &      &    &       &0.92550\\
&5& 0 & 0 &      &$50{^+},$ 0 & 9084.&10927 &      &      &    &       &0.99494\\
&4& 0 & 1 &      &$50{^-},$ 0 & 9138.&08249 &      &      &    &       &0.93036\\
&3& 0 & 2 &      &$41{^+},$ 0 & 9212.&94838 &      &      &    &       &0.97275\\
&2& 0 & 3 &      &$41{^-},$ 0 & 9213.&22358 &      &      &    &       &0.98537\\
&1& 0 & 4 &      &$32{^+},$ 0 & 9297.&77574 &      &      &    &       &0.97546\\
&0& 0 & 5 &      &$32{^-},$ 0 & 9335.&91658 &      &      &    &       &0.98839\\
& &   &   &      &            &      &     &      &      &     &      &         \\
&0&11 & 0 &$P$=11&$00{^+},$11 & 9070.&57613 &      &      &    &       &0.99780\\
&1& 9 & 0 &      &$10{^+},$ 9 & 9291.&71508 &      &      &    &       &0.99326\\
&0& 9 & 1 &      &$10{^-},$ 9 & 9292.&46814 &      &      &    &       &0.99761\\
&2& 7 & 0 &      &$20{^+},$ 7 & 9466.&61462 &      &      &    &       &0.96501\\
&1& 7 & 1 &      &$20{^-},$ 7 & 9469.&72515 &      &      &    &       &0.99400\\
&1& 7 & 2 &      &$11{^+},$ 7 & 9521.&24628 &      &      &    &       &0.96859\\
\hline
\end{tabular}
\end{center}
\end{table*}
\addtocounter{table}{-1}
\begin{table*}[tpb]
\caption{(cont.)}
 \small
\begin{center}
\begin{tabular}
%{cccc@{\hspace{2\tabcolsep}}c@{\hspace{2\tabcolsep}}r@{}l@{\hspace{2\tabcolsep}}r@{}l@{\hspace{2\tabcolsep}}r@{}l@{\hspace{2\tabcolsep}}c}\\
{ccccccr@{}lr@{}lr@{}lc}\\

\hline
&\multicolumn{3}{c}{Normal} &Polyad&    Local      & \multicolumn{2}{c}{E$_{cal}$} & \multicolumn{2}{c}{E$_{obs}$} & \multicolumn{2}{c}{E$_{obs.}$-E$_{cal.}$}&\%init.ket\\
\vspace{0.1cm}
&$\nu_1$ & $\nu_2$ &$\nu_3$ &$P$& $n_1n_2\pm,n_4$ & \multicolumn{2}{c}{$\mbox{cm}^{-1}$} & \multicolumn{2}{c}{$\mbox{cm}^{-1}$} & \multicolumn{2}{c}{$\mbox{cm}^{-1}$}  &$(Modulus)$ \\

\hline
&3& 5 & 0 &      &$30{^+},$ 5 & 9619.&75939 &      &      &    &       &0.98662\\
&2& 5 & 1 &      &$30{^-},$ 5 & 9619.&92503 &      &      &    &       &0.98828\\
&1& 5 & 2 &      &$21{^+},$ 5 & 9716.&46600 &      &      &    &       &0.98766\\
&0& 5 & 3 &      &$21{^-},$ 5 & 9731.&60237 &      &      &    &       &0.99318\\
&4& 3 & 0 &      &$40{^+},$ 3 & 9738.&90162 &      &      &    &       &0.98995\\
&3& 3 & 1 &      &$40{^-},$ 3 & 9738.&92598 &      &      &    &       &0.98947\\
&2& 3 & 2 &      &$31{^+},$ 3 & 9825.&56997 &      &      &    &       &0.99479\\
&1& 3 & 3 &      &$31{^-},$ 3 & 9825.&57090 &      &      &    &       &0.99480\\
&0& 3 & 4 &      &$22{^+},$ 3 & 9877.&91234 &      &      &    &       &0.92449\\
&5& 1 & 0 &      &$50{^+},$ 1 & 9886.&46046 &      &      &    &       &0.99379\\
&4& 1 & 1 &      &$50{^-},$ 1 & 9942.&38132 &      &      &    &       &0.93010\\
&3& 1 & 2 &      &$41{^+},$ 1 &10017.&42136 &      &      &    &       &0.97245\\
&2& 1 & 3 &      &$41{^-},$ 1 &10018.&67330 &      &      &    &       &0.98488\\
&1& 1 & 4 &      &$32{^+},$ 1 &10102.&34235 &      &      &    &       &0.97509\\
&0& 1 & 5 &      &$32{^-},$ 1 &10139.&21647 &      &      &    &       &0.98840\\
& &   &   &      &            &      &     &      &      &     &      &         \\
&0&12 & 0 &$P$=12&$00{^+},$12  & 9857.&90423 &      &      &    &       &0.99756\\
&1&10 & 0 &      &$10{^+},$10  &10081.&69213 &      &      &    &       &0.99722\\
&0&10 & 1 &      &$10{^-},$10  &10081.&93920 &      &      &    &       &0.99224\\
&2& 8 & 0 &      &$20 {^+},$ 8 &10257.&73493 &      &      &    &       &0.96378\\
&1& 8 & 1 &      &$20 {^-},$ 8 &10260.&88078 &      &      &    &       &0.99275\\
&0& 8 & 2 &      &$11 {^+},$ 8 &10312.&39973 &      &      &    &       &0.96809\\
&3& 6 & 0 &      &$30 {^+},$ 6 &10412.&83497 &      &      &    &       &0.98465\\
&2& 6 & 1 &      &$30 {^-},$ 6 &10413.&02778 &      &      &    &       &0.98612\\
&1& 6 & 2 &      &$21 {^+},$ 6 &10511.&49437 &      &      &    &       &0.98597\\
&0& 6 & 3 &      &$21 {^-},$ 6 &10524.&68974 &      &      &    &       &0.99270\\
&4& 4 & 0 &      &$40 {^+},$ 4 &10533.&98401 &      &      &    &       &0.98648\\
&3& 4 & 1 &      &$40 {^-},$ 4 &10534.&01023 &      &      &    &       &0.98612\\
\hline
\end{tabular}
\end{center}
\end{table*}
\addtocounter{table}{-1}
\begin{table*}[tpb]
\caption{(cont.)}
 \small
\begin{center}
\begin{tabular}
%{cccc@{\hspace{2\tabcolsep}}c@{\hspace{2\tabcolsep}}r@{}l@{\hspace{2\tabcolsep}}r@{}l@{\hspace{2\tabcolsep}}r@{}l@{\hspace{2\tabcolsep}}c}\\
{ccccccr@{}lr@{}lr@{}lc}\\

\hline
&\multicolumn{3}{c}{Normal} &Polyad&    Local      & \multicolumn{2}{c}{E$_{cal}$} & \multicolumn{2}{c}{E$_{obs}$} & \multicolumn{2}{c}{E$_{obs.}$-E$_{cal.}$}&\%init.ket\\
\vspace{0.1cm}
&$\nu_1$ & $\nu_2$ &$\nu_3$ &$P$& $n_1n_2\pm,n_4$ & \multicolumn{2}{c}{$\mbox{cm}^{-1}$} & \multicolumn{2}{c}{$\mbox{cm}^{-1}$} & \multicolumn{2}{c}{$\mbox{cm}^{-1}$}  &$(Modulus)$ \\
\hline

&2& 4 & 2 &      &$31 {^+},$ 4 &10622.&75025 &      &      &    &       &0.99038\\
&1& 4 & 3 &      &$31 {^-},$ 4 &10622.&75101 &      &      &    &       &0.99038\\
&0& 4 & 4 &      &$22 {^+},$ 4 &10673.&06131 &      &      &    &       &0.92314\\
&5& 2 & 0 &      &$50 {^+},$ 2 &10678.&75665 &      &      &    &       &0.99554\\
&4& 2 & 1 &      &$50 {^-},$ 2 &10678.&75667 &      &      &    &       &0.99555\\
&3& 2 & 2 &      &$41 {^+},$ 2 &10681.&56403 &      &      &    &       &0.99235\\
&2& 2 & 3 &      &$41 {^-},$ 2 &10737.&44623 &      &      &    &       &0.92971\\
&1& 2 & 4 &      &$32 {^+},$ 2 &10814.&78255 &      &      &    &       &0.97326\\
&0& 2 & 5 &      &$32 {^-},$ 2 &10815.&83798 &      &      &    &       &0.98356\\
&6& 0 & 0 &      &$60 {^+},$ 0 &10903.&29974 &      &      &    &       &0.97601\\
&5& 0 & 1 &      &$60 {^-},$ 0 &10920.&19816 &      &      &    &       &0.98695\\
&4& 0 & 2 &      &$51 {^+},$ 0 &10920.&29123 &      &      &    &       &0.98798\\
&3& 0 & 3 &      &$51 {^-},$ 0 &10936.&26317 &      &      &    &       &0.98808\\
&2& 0 & 4 &      &$42 {^+},$ 0 &11054.&52211 &      &      &    &       &0.88172\\
&1& 0 & 5 &      &$42 {^-},$ 0 &11070.&26392 &      &      &    &       &0.99042\\
&0& 0 & 6 &      &$33 {^+},$ 0 &11131.&78824 &      &      &    &       &0.89206\\

\hline
\end{tabular}
\end{center}
\end{table*}
\section{Conclusion}
We developed a formalism which allows a complete description of vibrational modes in local $XY_2$ type molecules. Stretching,
bending, interaction and transition operators have been built and analytical expressions for their matrix elements established in
the chains $u(2)\supset su(2)\supset so(2)$ and  $u(3)\supset u(2) \supset su(2) \supset so(2)$. Next a full symmetry adaptation
in the $C_{2v}$ molecular point group has been performed. This formalism has been applied to the $D_2S$ molecule where the 2:1
resonance between stretching and bending modes has been taken into account through an adapted Fermi-type operator the properties
of which have been discussed. From a simplified model we derived reasonable values for the highest stretching $N_s$ and bending
$N_b$ quantum numbers. Experimental data are reproduced with a standard deviation $\sigma=0.5 \mbox { cm}^{-1}$ and only 8
effective spectroscopic parameters. The dissociation energy calculated  with these parameters is in good agreement with the
experimental one and this also confirms our method for the determination of the $N_s$ value. Our approach will be applied to
other $XY_2$ molecular systems in a next paper.
\section{Acknowledgments}
This work was partially supported by a first financial support of the CNRS and the FRBR through a PICS (N$^o$ 170908) between the
Institut Carnot de Bourgogne - Universit\'{e} de Bourgogne - France and the Laboratory of Molecular Spectroscopy - Tomsk State
University - Russia and a second financial support through a grant (YS Fellowship N$^o$ 06-1000016-5751) INTAS - Bruxelles -
Belgium for Olga Gromova during her stays in the ICB. We want particularly to thank Martine Bonin from INTAS for all her nice
helps and excellent explanations.

%%%%%%%%%%%%%%%%%%%%%%%%%%%%%%%%%%%%%%%%%%%%%%%%%%%%%%%%%%%%%%%%%
%----------------------Appendix A---------------------------------
%%%%%%%%%%%%%%%%%%%%%%%%%%%%%%%%%%%%%%%%%%%%%%%%%%%%%%%%%%%%%%%%%%
%  set counter to zero and adapt the numbering to the section
\setcounter{equation}{0} \numberwithin{equation}{section}
\appendix
\section{Symmetry adapted Clebsch-Gordan coefficients and matrix elements for $\nu_2$}\label{appa}
The expressions for the ${~}^{[m_{4}\,-m_{4}]} G$ matrix elements are obtained from equations (\ref{eqb.19}, \ref{eqb.30}) which
lead to:
\begin{equation}\label{eqb.34}
\begin{split}
{~}^{[m'_{4}\,-m'_{4}]} G_{0A_1}^{m}&= i^{m'_4} \delta_{m,0}, \\
{~}^{[m'_{4}\,-m'_{4}]} G_{|m| \varepsilon A_1}^{m}&=\frac{i^{\varepsilon}}{\sqrt{2}}\delta_{m,|m|},\\
{~}^{[m'_{4}\,-m'_{4}]} G_{|m| \varepsilon
A_1}^{m}&=\frac{i^{\varepsilon}}{\sqrt{2}}(-1)^{\varepsilon}(-1)^{m'_4+m}\delta_{m,-|m|}.
\end{split}
\end{equation}
As a result:\\
$\bullet$ For $r=0$ we obtain the CG (\ref{eqb.33})
\begin{eqnarray}\label{eqb.35}
\lefteqn{F\begin{array}{ccc} 0\,A_1 & m' A_1 & ([N_b\,0]J_b) \\
([m'_{4}\,-m'_{4}]j_b & [N_b\,0]J_b ) & m'' A_1
\end{array}= } \nonumber \\&&=i^{-m'_4} \,C\begin{array}{ccc} 0 & m' &(J_b) \\
(j_b & J_b)& m''\end{array}=i^{-m'_4} \,\delta_{m',m''},
\end{eqnarray}
and for the matrix elements (\ref{eqb.32}):
\begin{eqnarray}\label{eqb.36}
\lefteqn{\langle\langle \lbrack N_b\,0\rbrack J_b\, m'' A_1 |{}^{[m_4\,-m_4]}\mathcal{B}^{(j_b)}_{0 A_1} |\lbrack N_b\,0\rbrack
J_b \,m' A_1 \rangle\rangle =i^{m'_4} \delta_{m',m''}} \nonumber \\&&\!\times (2J_b+1)^{-\frac{1}{2}}
\left([N_b\,0]J_b||{}^{[m_4\,-m_4]}\mathcal{B}^{(j_b)}{}||[N_b\,0]J_b\right).
\end{eqnarray}
$\bullet$ For $r=|m| \varepsilon$. With (\ref{eqb.34}) we obtain in this case for the CG (\ref{eqb.33})
\begin{eqnarray}\label{eqb.37}
\lefteqn{ F\begin{array}{ccc} |m| \varepsilon A_1 & m' A_1 & ([N_b\,0]J_b) \\
([m'_{4}\,-m'_{4}]j_b & [N_b\,0]J_b ) & m'' A_1
\end{array} = }\nonumber \\
&&~~~~\frac{(-i)^{\varepsilon}}{\sqrt{2}}\left( \,C\begin{array}{ccc}~ m & m' &(J_b) \\
(j_b & J_b)& m''\end{array} \right.\nonumber \\&&~~~~\left. +(-1)^{\varepsilon}(-1)^{m'_4+m}\ \,C\begin{array}{ccc} -m & m' &(J_b) \\
(~j_b & J_b)& m''\end{array} \right).
\end{eqnarray}
This last expression is to be taken with $m>0$ (one can also set $|m|$ in the CG). We then have for the matrix elements
(\ref{eqb.32}):
\begin{eqnarray}\label{eqb.38}
\lefteqn{ \langle\langle \lbrack N_b\,0\rbrack J_b\, m'' A_1 |{}^{[m_4\,-m_4]}\mathcal{B}^{(j_b)}_{|m| \varepsilon A_1} |\lbrack
N_b\,0\rbrack J_b \,m' A_1 \rangle\rangle =} \nonumber \\
 & & (2J_b+1)^{-\frac{1}{2}}\frac{i^{\varepsilon}}{\sqrt{2}}\left( \,C\begin{array}{ccc}~ m & m' &(J_b) \\
(j_b & J_b)& m''\end{array} \right.\nonumber \\&& \left.+(-1)^{\varepsilon}(-1)^{m'_4+m}\ \,C\begin{array}{ccc} -m & m' &(J_b) \\
(~j_b & J_b)& m''\end{array} \right)
\nonumber \\
& & \nonumber \\ & & \times \left([N_b\,0]J_b||{}^{[m_4\,-m_4]}\mathcal{B}^{(j_b)}{}||[N_b\,0]J_b\right),
\end{eqnarray}
where we took into account that the $su(2)$ standard CG are real.
%%%%%%%%%%%%%%%%%%%%%%%%%%%%%%%%%%%%%%%%%%%%%%%%%%%%%%%%%%%%%%%%%
%----------------------Appendix B---------------------------------
%%%%%%%%%%%%%%%%%%%%%%%%%%%%%%%%%%%%%%%%%%%%%%%%%%%%%%%%%%%%%%%%%%
\section{Symmetry adapted Clebsch-Gordan coefficients for $\nu_1$, $\nu_3$}\label{appb}
We give below the symmetry adapted CG coefficients used in the study of stretching modes. They are obtained from equations
(\ref{eqs.24a}, \ref{eqs.24b}, \ref{eqs.29}). The remaining standard $su(2)$ coefficients are taken from \cite{EDM1974}.\\
$\bullet$ $|m|=|m'|=|m''|=0$\\This condition implies $n'_s=2j'_s$, $n''_s=2j''_s$ and $m_1+m_2$ even.
\begin{eqnarray}\label{eqs.30}
\lefteqn{F\begin{array}{ccc} 0 \Gamma& 0\,A_1 & ([n_s''\,0]j_s'') \\
([m_{1}\,-m_{2}]j & [n_s'\,0]j_s' ) & 0\,A_1
\end{array} = }\nonumber \\
& &\! i^{-m_2-j}\, C\begin{array}{ccc} 0 & 0 & (j''_s) \\
(j & j'_s ) & 0 \end{array}\!=i^{-j}(n'_s+m_1-m_2+1)^{\frac{1}{2}}  \nonumber \\
&&\times \left[\frac{(n'_s-m_2)!\,(m_1)!\,(m_2)!}{(n'_s+m_1+1)!}\right]^{\frac{1}{2}}
\nonumber \\
&& \times \frac{((n'_s+m_1)/2)!}{((n'_s-m_2)/2)!\,(m_1/2)!\,(m_2/2)!}\,\delta_{\Gamma,A_1}.
\end{eqnarray}
As it is known the CG on the second line in non zero only if $j+j'_s+j''_s$ is even which corresponds to the selection rule
$\Gamma=A_1$.\\
$\bullet$ $|m|=|m'|=0$, $|m''|\neq0$ or $|m|=|m''|=0$, $|m'|\neq0$ or $|m'|=|m''|=0$, $|m|\neq0$\\
It is easily checked that all $F$ coefficients are zero in these cases.\\
$\bullet$ $|m|=0$, $|m'|\neq 0$, $|m''|\neq0$
\begin{eqnarray}\label{eqs.33}
F\begin{array}{ccc} 0 \Gamma& |m'| \Gamma' & ([n_s''\,0]j_s'') \\
([m_{1}\,-m_{2}]j & [n_s'\,0]j_s' ) & |m''| \Gamma''
\end{array} =\,i^{-m_2}\,i^{-j-\varepsilon}\nonumber\\
 \times \, C\begin{array}{ccc} 0 & m' & (j''_s) \\
(j & j'_s ) & m' \end{array}\left(\frac{1+(-1)^{m_1+\varepsilon'+\varepsilon''}}{2}\right)\delta_{|m'|,|m''|}
\end{eqnarray}
We note that the factor $1+(-1)^{m_1+\varepsilon'+\varepsilon''}$ just traduces the selection rule $\Gamma \times
\Gamma'=\Gamma''$. In fact for the CG to be non zero we must have $(-1)^{m_1+\varepsilon'+\varepsilon''}=1$, that is for $m_1$
even $\Gamma=A_1$ and $\varepsilon'=\varepsilon''$ that is $\Gamma'=\Gamma''$ ($A_1$ or $B_1$) ; for $m_1$ odd, $\Gamma=B_1$ and
$\varepsilon'\neq\varepsilon''$ that is $\Gamma'\neq \Gamma''$ ($(\Gamma',\Gamma'')=(A_1,B_1)$ or $(\Gamma',\Gamma'')=(B_1,A_1)$).\\
$\bullet$ $|m'|=0$, $|m|\neq 0$, $|m''|\neq0$\\In this case we necessarily have $j'_s$ integer so $n'_s$ even and $\Gamma'=A_1$.
Here again a factor $1+(-1)^{\varepsilon+\varepsilon''}$ appears in the computation, which is equivalent to $\Gamma=\Gamma''$. We
find
\begin{eqnarray}\label{eqs.34}
\lefteqn{F\begin{array}{ccc} |m| \Gamma& 0\, A_1 & ([n_s''\,0]j_s'') \\
([m_{1}\,-m_{2}]j & [n_s'\,0]j_s' ) & |m''| \Gamma''
\end{array} = \,i^{-m_2}\,i^{-j-|m|-\varepsilon} } \nonumber\\
&&\times \ C\begin{array}{ccc} m & 0 & (j''_s) \\
(j & j'_s ) & m \end{array}\,\delta_{|m|,|m''|}\,\delta_{\Gamma,\Gamma''}.
\end{eqnarray}
$\bullet$ $|m''|=0$, $|m|\neq 0$, $|m'|\neq0$.\\
This case is similar to the preceding one. We necessarily have $j''_s$ integer so $n''_s$ even and $\Gamma''=A_1$. Here again a
factor $1+(-1)^{\varepsilon+\varepsilon'}$ appears in the computation, which is equivalent to $\Gamma=\Gamma'$. We obtain
\begin{eqnarray}\label{eqs.35}
\lefteqn{F\begin{array}{ccc} |m| \Gamma& |m'| \Gamma' & ([n_s''\,0]j_s'') \\
([m_{1}\,-m_{2}]j & [n_s'\,0]j_s' ) & 0\, A_1
\end{array} = \,i^{m_1}\,i^{j-|m|+\varepsilon} }\nonumber\\
&&\times \ C\begin{array}{ccc} m & -m & (j''_s) \\
(j & j'_s ) & 0 \end{array}\,\delta_{|m|,|m'|}\,\delta_{\Gamma,\Gamma'}.
\end{eqnarray}
We underline that in equations (\ref{eqs.33}, \ref{eqs.34}, \ref{eqs.35}) the $su(2)$ CG in the right member is to be taken with
$m$ (or $m'$) positive.\\
$\bullet$ $|m'|\neq 0$, $|m'|\neq 0$, $|m''|\neq0$.\\
In all cases a coefficient $1+(-1)^{\varepsilon+\varepsilon'+\varepsilon''}$ appears in the calculation, which traduces the
selection rule $\Gamma''=\Gamma \times \Gamma'$. We obtain the following non-zero coefficients:\\
- For $|m''|=|m|+|m'|$
\begin{eqnarray}\label{eqs.36}
\lefteqn{F\begin{array}{ccc} |m| \Gamma& |m'| \Gamma' & ([n_s''\,0]j_s'') \\
([m_{1}\,-m_{2}]j & [n_s'\,0]j_s' ) & |m''| \Gamma''
\end{array}  = \,i^{-m_2}\,i^{-j-|m|-\varepsilon} } \nonumber\\
&\times  \displaystyle{\frac{1}{\sqrt{2}}}\ C\begin{array}{ccc} m & m' & (j''_s) \\
(j & j'_s ) & m+m' \end{array}\,\delta_{|m''|,|m|+|m'|}\,\delta_{\Gamma'',\Gamma\times\Gamma'}.
\end{eqnarray}
- For $m > m' >0$
\begin{eqnarray}\label{eqs.37}
\lefteqn{F\begin{array}{ccc} |m| \Gamma& |m'| \Gamma' & ([n_s''\,0]j_s'') \\
([m_{1}\,-m_{2}]j & [n_s'\,0]j_s' ) & |m''| \Gamma''
\end{array} \! = i^{n''+\varepsilon''}\,i^{j-|m|+\varepsilon} }\nonumber\\
&\!\times \displaystyle{\frac{1}{\sqrt{2}}}\, C\begin{array}{ccc} m & -m' & (j''_s) \\
(j & j'_s ) & m-m' \end{array}\!\delta_{|m''|,|m|-|m'|}\,\delta_{\Gamma'',\Gamma\times\Gamma'}.
\end{eqnarray}
- For $m' > m >0$
\begin{eqnarray}\label{eqs.38}
\lefteqn{F\begin{array}{ccc} |m| \Gamma& |m'| \Gamma' & ([n_s''\,0]j_s'') \\
([m_{1}\,-m_{2}]j & [n_s'\,0]j_s' ) & |m''| \Gamma''
\end{array}  = \,i^{m_1}\,i^{j-|m|+\varepsilon} }\nonumber\\
&\!\times   \displaystyle{\frac{1}{\sqrt{2}}}\, C\begin{array}{ccc} m & -m' & (j''_s) \\
(j & j'_s ) & m-m' \end{array}\!\delta_{|m''|,|m'|-|m|}\,\delta_{\Gamma'',\Gamma\times\Gamma'}.
\end{eqnarray}


\begin{thebibliography}{00}

%%\bibitem{}
\bibitem{LAM2008} J. Lamouroux, L. R\'{e}galia-Jarlot, Vl.G. Tyuterev, X. Thomas, P. Von der Heyden, S.A. Tashkun, Yu. Borkov, J. Mol. Spectrosc. 250 (2008) 117-125.
\bibitem{ZEL2008} Z. Zelinger, A. Perrin, M. St\v{r}i\v{z}\'{\i}k, P. Kub\'{a}t, J. Mol. Spectrosc. 249 (2008), 117-120.
\bibitem{FOR2007} T.A. Ford, J. Mol. Struct. 834-836 (2007) 30-41.
\bibitem{FRI2005} A.J. Friedson, Icarus, 177 (2005) 1-17.
\bibitem{SMI2001} M. Smirnov, PhD thesis, IAO-Tomsk, Russie, LPM-Orsay, France 2001.
\bibitem{MIN1989} Y.C. Minh, W.M. Irvine, L. M. Ziurys, Astrophys. J. Lett. 345 (1989) L63-L66.
\bibitem{HER1989} E. Herbst, D.J. Defrees, W. Koch, Notices of the Roy. Astronomic. Soc. 237 (1989) 1057-1065.
\bibitem{GRI1987} R.J.A. Grim, J.M. Greenberg, Astron. Astrophys. 181 (1987) 155-168.
\bibitem{PLI1967} J. Pliva, V. \v{S}pirko, D. Papou\v{s}ek, J. Mol. Spectrosc. 23 (1975) 331-342.
\bibitem{COO1975} R.L. Cook, F.C. De Lucia, P. Helminger, J. Mol. Struct. 28 (1967) 237-246.
\bibitem{GIL1985} J.R. Gillis, R.D. Blatherwick, F.S. Bonomo, J. Mol. Spectrosc. 114 (1985) 228-233.
\bibitem{CAM1988} C. Camy-Peyret, J.M. Flaud, A. N'Gom, Mol. Phys. 65 (1988) 649-657.
\bibitem{KOZ1994} I.N. Kozin, P. Jensen, J. Mol. Spectrosc. 163 (1994) 483-509.
\bibitem{LIU2006} A.W. Liu, O.N. Ulenikov, G.A. Onopenko, O.V. Gromova, E.S. Bekhtereva. L. Wan, L.Y. Wan, S.M. Hu, J.M. Flaud, J. Mol. Spectrosc. 238 (2006) 23-40.
\bibitem{HAL1987} L. Halonen, T. Carrington Jr, J. Chem. Phys. 88 (1988) 4171-4185.
\bibitem{IAC1990} F. Iachello, S. Oss, J. Mol. Spectrosc. 142 (1990) 85-107.
\bibitem{ZHE2000} Y. Zheng, S. Ding, J. Mol. Spectrosc. 201 (2000) 109-115.
\bibitem{NAU2001} O. Naumenko, A. Campargue, J. Mol. Spectrosc. 210 (2001) 224-252.
\bibitem{TYU2001} Vl.G. Tyuterev, Chem. Phys. Lett. 348 (2001) 223-234.
\bibitem{TYU2004} Vl.G. Tyuterev, L. Regalia-Jarlot,  D.W. Schwenke, S.A. Tashkun, Y.G. Borkov, C. R. Phys. 5 (2004)189-199.
\bibitem{MIC1987} F. Michelot, J. Moret-Bailly, J. Physique 48 (1987) 51-72.
\bibitem{LER1992} C. Leroy, F. Michelot, J. Mol. Spectrosc. 151 (1992) 71-96.
\bibitem{LER1994} C. Leroy, F. Michelot, Can. J. Phys. 72 (1994) 274-289.
\bibitem{MIC2004} F. Michelot, M. Rey, Eur. Phys. J. D, 30 (2004) 181-189.
\bibitem{MIC2005} F. Michelot, M. Rey, Eur. Phys. J. D, 33  (2005) 357-386.
\bibitem{MIC2007} F. Michelot, M. Rey, Eur. Phys. J. D, 44  (2007) 467-495.
\bibitem{WYB1974} B.G. Wybourne, Classical Lie Groups for Physicists, Wiley-Interscience, New York, 1974.
\bibitem{GEL1950} I.M. Gel'fand, M.L. Zetlin, Dokl. Akad. Nauk. (in Russian) 71 (1950) 825-828.
\bibitem{LOU1970} J. Louck, Amer. J. Phys., 38 (1970) 3-42.
\bibitem{EDM1974} A.R. Edmonds,Angular Momentum in Quantum Mechanics, Princeton University Press, 1974.
\bibitem{HOL1971}
W. Holmann III, L.C. Biedenharn, in Group Theory and Its Applications,edited by E.M. Loebl, Academic Press, New York, 1971 Vol.
II, p.1-73.
\bibitem{LER1991} C. Leroy, PhD thesis, University of Dijon, France 1991.
\bibitem{BOU1996} V. Boujut, PhD thesis, University of Dijon, France 1996.
\bibitem{PLU2005} L. Pluchart, C. Leroy, N. Sanzharov, F. Michelot, E.S. Bekhtereva, O.N. Ulenikov, J. Mol. Spectrosc. 232
(2005) 119-136.
\bibitem{SAN2008} N.A. Sanzharov, C. Leroy, O.N. Ulenikov, E.S. Bekhtereva, J. Mol. Spectrosc,  247 (2008) 1-24.
\bibitem{PEE2002} L.R. Peebles, P. Marshall, J. Chem. Phys. 117 (2002) 3132-3138.
\end{thebibliography}
\end{document}